\documentstyle[11pt]{article}
\newcounter{subequation}[equation]

\makeatletter 
 
\def\thesubequation{\theequation\@alph\c@subequation}
\def\@subeqnnum{{\rm (\thesubequation)}}
\def\slabel#1{\@bsphack\if@filesw {\let\thepage\relax
   \xdef\@gtempa{\write\@auxout{\string
      \newlabel{#1}{{\thesubequation}{\thepage}}}}}\@gtempa
   \if@nobreak \ifvmode\nobreak\fi\fi\fi\@esphack}
\def\subeqnarray{\stepcounter{equation}
\let\@currentlabel=\theequation\global\c@subequation\@ne
\global\@eqnswtrue
\global\@eqcnt\z@\tabskip\@centering\let\\=\@subeqncr
$$\halign to \displaywidth\bgroup\@eqnsel\hskip\@centering
  $\displaystyle\tabskip\z@{##}$&\global\@eqcnt\@ne
  \hskip 2\arraycolsep \hfil${##}$\hfil
  &\global\@eqcnt\tw@ \hskip 2\arraycolsep
  $\displaystyle\tabskip\z@{##}$\hfil
   \tabskip\@centering&\llap{##}\tabskip\z@\cr}
\def\endsubeqnarray{\@@subeqncr\egroup
                     $$\global\@ignoretrue}
\def\@subeqncr{{\ifnum0=`}\fi\@ifstar{\global\@eqpen\@M
    \@ysubeqncr}{\global\@eqpen\interdisplaylinepenalty \@ysubeqncr}}
\def\@ysubeqncr{\@ifnextchar [{\@xsubeqncr}{\@xsubeqncr[\z@]}}
\def\@xsubeqncr[#1]{\ifnum0=`{\fi}\@@subeqncr
   \noalign{\penalty\@eqpen\vskip\jot\vskip #1\relax}}
\def\@@subeqncr{\let\@tempa\relax
    \ifcase\@eqcnt \def\@tempa{& & &}\or \def\@tempa{& &}
      \else \def\@tempa{&}\fi
     \@tempa \if@eqnsw\@subeqnnum\refstepcounter{subequation}\fi
     \global\@eqnswtrue\global\@eqcnt\z@\cr}
\let\@ssubeqncr=\@subeqncr
\@namedef{subeqnarray*}{\def\@subeqncr{\nonumber\@ssubeqncr}\subeqnarray}
\@namedef{endsubeqnarray*}{\global\advance\c@equation\m@ne%
                           \nonumber\endsubeqnarray}

\makeatletter
\@addtoreset{equation}{section}
\makeatother
\renewcommand{\theequation}{\thesection.\arabic{equation}}

\def\dalemb#1#2{{\vbox{\hrule height .#2pt
        \hbox{\vrule width.#2pt height#1pt \kern#1pt
                \vrule width.#2pt}
        \hrule height.#2pt}}}

\let\a=\alpha    \let\e=\epsilon
  \let\q=\theta  \let\k=\kappa
 \let\m=\mu \let\n=\nu

\def\nn{\nonumber} \def\bd{\begin{document}} \def\ed{\end{document}}
\def\ds{\documentstyle} \let\fr=\frac \let\bl=\bigl \let\br=\bigr
\let\Br=\Bigr \let\Bl=\Bigl 
\let\bm=\bibitem
\let\na=\nabla
\let\pa=\partial \let\ov=\overline
\def\ie{{\it i.e.\ }} 
\newcommand{\be}{\begin{equation}} 
\newcommand{\ee}{\end{equation}} 
\def\ba{\begin{array}}
\def\ea{\end{array}}
\def\ft#1#2{{\textstyle{{\scriptstyle #1}\over {\scriptstyle #2}}}}
\def\fft#1#2{{#1 \over #2}}
\def\del{\partial}
\def\sst#1{{\scriptscriptstyle #1}}
\def\oneone{\rlap 1\mkern4mu{\rm l}}
\def\e7{E_{7(+7)}}
\def\td{\tilde}
\def\wtd{\widetilde}
\def\im{{\rm i}}
\def\bog{Bogomol'nyi\ }
\def\q{{\tilde q}}
\def\hast{{\hat\ast}}
\def\0{{\sst{(0)}}}
\def\1{{\sst{(1)}}}
\def\2{{\sst{(2)}}}
\def\3{{\sst{(3)}}}
\def\4{{\sst{(4)}}}
\def\5{{\sst{(5)}}}
\def\6{{\sst{(6)}}}
\def\7{{\sst{(7)}}}
\def\8{{\sst{(8)}}}
\def\n{{\sst{(n)}}}
\def\oo{{\"o}}
\def\hA{\hat{\cal A}}
\def\ns{{\sst {\rm NS}}}
\def\rr{{\sst {\rm RR}}}
\def\tH{{\widetilde H}}
\def\tB{{\widetilde B}}
\def\cA{{\cal A}}
\def\cF{{\cal F}}
\def\tF{{\wtd F}}
\def\Z{\rlap{\sf Z}\mkern3mu{\sf Z}}
\def\ep{{\epsilon}}
\def\IIA{{\rm IIA}}
\def\IIB{{\rm IIB}}
\def\ads{{\rm AdS}}
\def\R{\rlap{\rm I}\mkern3mu{\rm R}}

\newcommand{\ho}[1]{$\, ^{#1}$}
\newcommand{\hoch}[1]{$\, ^{#1}$}
\newcommand{\bea}{\begin{eqnarray}} 
\newcommand{\eea}{\end{eqnarray}} 
\newcommand{\ra}{\rightarrow}
\newcommand{\lra}{\longrightarrow}
\newcommand{\Lra}{\Leftrightarrow}
\newcommand{\ap}{\alpha^\prime}
\newcommand{\bp}{\tilde \beta^\prime}
\newcommand{\tr}{{\rm tr} }
\newcommand{\Tr}{{\rm Tr} } 
\newcommand{\NP}{Nucl. Phys. }
\newcommand{\tamphys}{\it Center for Theoretical Physics,
Texas A\&M University, College Station, TX 77843}
\newcommand{\upenn}{\it Dept. of Physics and Astronomy, 
University of Pennsylvania,
Philadelphia, PA 19104}

\newcommand{\auth}{M. Cveti\v{c}\hoch{\dagger1}, H. L\"u\hoch{\dagger1} and 
C.N. Pope\hoch{\ddagger2}}

\begin{document}
\begin{flushright}
\hfill{CTP TAMU-37/98 \\ 
UPR/0817-T \\
October 1998}\\
\hfill{\bf hep-th/9810123}\\
\end{flushright}


\begin{center}
{\large {\bf Spacetimes of Boosted $p$-branes, and CFT
in Infinite-momentum Frame}} 

\vspace{20pt}

\auth

\vspace{10pt}
{\hoch{\dagger}\upenn}

\vspace{10pt}
{\hoch{\ddagger}\tamphys}

\vspace{30pt}

\underline{ABSTRACT}
\end{center}

     We study the spacetimes of the near-horizon regions in D3-brane,
M2-brane and M5-brane configurations, in cases where there is a
pp-wave propagating along a direction in the world-volume. While
non-extremal configurations of this kind locally have the same
Carter-Novotn\'y-Horsk\'y-type metrics as those without the wave,
taking the BPS limit results instead in Kaigorodov-type metrics, which
are homogeneous, but preserve $\ft14$ of the supersymmetry, and have
global and local structures that are quite different from the
corresponding anti-de Sitter spacetimes associated with solutions
where there is no pp-wave. We show that the momentum density of the
system is non-vanishing and held fixed under the gravity decoupling
limit.  In view of the AdS/CFT correspondence, M-theory and type IIB
theory in the near-horizon region of these boosted BPS-configurations
specifies the corresponding CFT on the boundary in an
infinitely-boosted frame with constant momentum density.  We model the
microstates of such boosted configurations (which account for the
microscopic counting of near-extremal black holes in $D=7$, $D=9$ and
$D=6$) by those of a boosted dilute massless gas in a $d=4$, $d=3$ and
$d=6$ spacetime respectively.  Thus we obtain a simple description for
the entropy of 2-charge black holes in $D=7,9$ and 6 dimensions.  The
paper includes constructions of generalisations of the Kaigorodov and
Carter-Novotn\'y-Horsk\'y metrics in arbitrary spacetime dimensions,
and an investigation of their properties.

{\vfill\leftline{}\vfill
\vskip 10pt \footnoterule {\footnotesize \hoch{1} Research supported
in part by DOE grant DOE-FG02-95ER40893
\vskip  -12pt} \vskip   14pt
{\footnotesize
        \hoch{2}        Research supported in part by DOE 
grant DOE-FG03-95ER40917 \vskip -12pt}  \vskip  14pt
}

\pagebreak
\setcounter{page}{1}

\tableofcontents
\addtocontents{toc}{\protect\setcounter{tocdepth}{2}}
\newpage

\section{Introduction\label{sec:intro}}

One of the important implications of the non-perturbative aspects of
M-theory is the counting of microstates for near-BPS configurations
such as black holes and $p$-branes.  For near-BPS black holes in $D=5$
and $D=4$, this counting can be carried out precisely, both from the
D-brane perspective \cite{strom96a,calmal,horstr,ell,all} as well as
by the counting of the small-scale oscillations of the effective
string theory in the NS-NS sector~\cite{CvTsI,TsI,das,TsII,CvTsII}.
Interestingly, these examples reduce to the counting of the degrees of
freedom of an effective string theory, \ie a (1+1)-dimensional
conformal field theory (CFT). (These degrees of freedom are
effectively modelled by those of a dilute gas of massless particles in
1+1 dimensions.)  On the other hand, the entropy of the D3-brane,
M2-brane, and M5-brane, can be modelled (up to a prefactor) by a
dilute gas in $d=4$, $d=3$ and $d=6$ respectively \cite{gkp,kt,kt1}.

   Maldacena's conjecture that relates Type IIB string theory on
anti-de Sitter (AdS) spacetime to conformal field theory (CFT) on its
boundary (the ``AdS/CFT correspondence'') \cite{mald}, which has been
investigated by earlier works \cite{gkp,Klebanov,Gubser2}, has
initiated broad efforts to test it at the level of the spectrum and
correlation functions.  One of the important implications of the
conjecture is the fresh perspective that it sheds on the microscopics
of black holes.  It was observed in \cite{hyun,ss} that the black
holes in $D=5$ and $D=4$ are related to the three-dimensional BTZ
black hole \cite{btz} (see also \cite{hw}). This leads to a new
derivation of black hole entropy\cite{strom,dublinbtz}, by studying
the decoupling regime of the near-horizon black hole geometry.  The
central observation is that, when embedded in a higher-dimensional
space, the near-horizon geometry of black holes in $D=5$
\cite{strom,CvLaI} and $D=4$ \cite{BaLa,CvLaII} contains locally the
three-dimensional anti-de Sitter spacetime (AdS$_3$), whose quantum
states are described by a two-dimensional conformal field theory on
its asymptotic boundary \cite{BH}.  The counting of states in this CFT
is then used to reproduce the black hole entropy for near-extremal
static \cite{strom,BaLa} and rotating \cite{CvLaI,CvLaII} black holes
in $D=4$ and $D=5$ respectively.
    
      In this paper, we address a number of related issues: 
\bigskip

\noindent{$\bullet$} {\it Near-horizon geometry of boosted
p-branes/CFT in infinite-momentum frame}  

We address the AdS$_D$/CFT correspondence to cases where there is a
pp-wave propagating along a direction in the world-volume of the
classical $p$-brane configuration.  One has to distinguish two cases,
depending upon whether or not the configuration is BPS saturated.  In
the non-BPS case, the effect of the inclusion of the pp-wave is
locally equivalent to performing a Lorentz boost transformation along
the direction of propagation of the wave.  (If the direction along
which the pp-wave propagates is uncompactified, then the equivalence
is in fact valid globally, while if the direction is wrapped on a
circle, it is only valid locally.)  For this reason, $p$-branes with
superimposed pp-waves propagating on their world-volumes are often
referred to as {\it boosted} $p$-branes; one should bear in mind
though that the global structure may not be precisely describable by a
Lorentz boost.  In the case of BPS $p$-branes, on the other hand, the
inclusion of the pp-wave leads to a metric that is not even locally
equivalent to the one where there is no wave.  This is because in the
BPS limit the Lorentz boost that relates the two metrics becomes
singular, corresponding to a boost with velocity approaching the speed
of light.  Thus in the BPS limit one has two distinct configurations,
which are not even locally equivalent, corresponding to the cases with
and without the pp-wave. In this case, although the term ``boosted
$p$-brane'' is sometimes used, the expression is somewhat of a
misnomer.

  For the BPS D3-brane, M2-brane and M5-brane configurations where
there is no pp-wave, the near-horizon geometries (corresponding to the
decoupling limit) are those of AdS$_5\times S^5$, AdS$_4\times S^7$ or
AdS$_7\times S^4$ respectively \cite{dgt,ght}.  Equivalently, one can
think of performing a compactification on the 5-spheres, 7-spheres or
4-spheres that foliate the space transverse to the $p$-brane, in which
case, in the near-horizon regime, the corresponding AdS spacetimes
arise as solutions of the compactified theories.  On the other hand,
with the inclusion of a pp-wave propagating on the BPS $p$-brane we
find that the AdS metric is replaced by a new type of metric, which in
four dimensions was first constructed by Kaigorodov \cite{kaig}.  (The
four-dimensional metric is of type N in the Petrov classification.
See also discussions in \cite{siklos,grbook,pod}.) In this paper, we
construct arbitrary-dimensional generalisations of the Kaigorodov
metric, which include the $D=5,4$ and 7 cases arising in the
near-horizon regions of the boosted D3-brane, M2-brane and M5-brane.
Like AdS, these are homogeneous Einstein metrics, but they differ
significantly in both their local and global structures.  In
particular, although they approach AdS locally at infinity, their
boundaries are related to those of the AdS metrics by an infinite
Lorentz boost.  Thus one may say that the boundary of the generalised
Kaigorodov metric is in an infinite-momentum frame.  Furthermore, we
show that in the gravity decoupling limit, in order to maintain the
structure of the Kaigorodov metric, the momentum density (momentum per
unit $p$-volume) must be held fixed.  (The metric recovers the form of
the AdS spactime if instead the momentum density vanishes.) A
consequence of this generalisation of the spacetime is that the
boundary theory will now be a CFT with an infinite boost, but with a
constant momentum density.  This new correspondence implies that the
entropy of the near-extremal D3-brane or M-brane with pp-wave can be
modelled by a dilute massless gas in an appropriately-boosted frame of
the world-volume spacetime of the $p$-brane.  We show that this is
indeed the case, and that the contribution to the entropy in the
boosted case is precisely accounted for by the Lorentz contraction
factor $1/\gamma$ along the boost direction, implying that the entropy
density (entropy per unit $p$-volume) is enlarged by a factor of
$\gamma$.  (This observation has also been made in \cite{hm,dmrr}.)

    The situation is somewhat different in the case of non-extremal
$p$-branes.  We show that the spherical reductions of the
configurations with pp-waves give rise to inhomogeneous Einstein
metrics, which generalise the Carter-Novotn\'y-Horsk\'y metric
\cite{carter,novhor} of four dimensions.  We again construct
arbitrary-dimensional generalisations, which encompass the cases that
arise from the spherical reductions of the D3-brane, M2-brane and
M5-brane, and we study some of the pertinent properties of these
metrics.  As we noted above, in these non-extremal configurations
there is locally no distinction between the case where there is a
superimposed pp-wave, and the case with no pp-wave.  This is because a
coordinate transformation allows the harmonic function associated with
the pp-wave to be set to unity.  Consequently the local form of the
Carter-Novotn\'y-Horsk\'y metrics is the same whether or not a pp-wave
is included in the original $p$-brane solution.  The coordinate
transformation becomes singular in the extremal limit, which explains
why there are two distinct cases in the extremal situation, leading
either to the AdS or else to the generalised Kaigorodov metrics after
spherical reduction, but only the single case of the generalised
Carter-Novotn\'y-Horsk\'y metrics in the spherically-reduced
non-extremal situations.

    We may summarise the situation in the following Table.  If we
begin with a non-dilatonic $p$-brane in $\wtd D$ dimensions, and
perform a dimensional reduction on the foliating $(\wtd
D-p-2)$-spheres in the transverse space, then according to whether the
$p$-brane is extremal or non-extremal, and whether or not there is a
superimposed pp-wave, the lower-dimensional metric (of dimension
$n=p+2$) will be of the form:

\bigskip\bigskip

\begin{center}
\begin{tabular}{|c|c|c|}\hline
   & No Wave  & Wave  \\ \hline
Extremal & AdS$_n$  & K$_n$ \\ \hline
Non-extremal  & C$_n$ & C$_n$ \\ \hline
\end{tabular}
\end{center}

\bigskip

\centerline{Table 1: Spherical reductions of non-dilatonic $p$-branes}

\bigskip\bigskip

   Here, K$_n$ denotes the $n$-dimensional generalisation of the
Kaigorodov metric, obtained in Appendix A, and C$_n$ denotes the
$n$-dimensional generalisation of the Carter-Novotn\'y-Horsk\'y metric,
obtained in Appendix C.

    Using the fact that horizon area, and hence entropy, is preserved
under dimensional reduction, we show how the entropies of certain of
the black holes can be related to the entropies calculated in the
associated generalised Kaigorodov or Carter-Novotn\'y-Horsk\'y
metrics.  The results that we obtain in this paper are generalisations
of results obtained previously for the BTZ metrics.  In particular,
the extremal BTZ metric (where the angular momentum $J$ and mass $M$
are related by $J=M\, \ell$, where $-2\ell^{-2}$ is the cosmological
constant) is equivalent to K$_3$, the specialisation of the
generalised Kaigorodov metrics to the case $D=3$.  Likewise, the
non-extremal BTZ metric is equivalent to C$_3$, the specialisation of
the generalised Carter-Novotn\'y-Horsk\'y metrics to the case $D=3$.
\bigskip

\noindent{$\bullet$}{\it Black-hole microstate counting for $D>5$}.

    The above aspect of the AdS/CFT correspondence allows for a study
of the microscopics of general static near-extremal black holes in
$D=7$, 9 and 6.  In other words, if the entire spatial world-volume of
a near-extremal D3-brane, M2-brane or M5-brane configuration with a
pp-wave is compactified on a torus, we obtain a two-charge static
near-extremal black hole in $D=7$, $D=9$ or $D=6$ respectively.  (In
$6\le D\le 9$, such two-charge black holes are generating solutions
for the most general black holes of the toroidally compactified
heterotic and Type II string theories~\cite{lpsol,CvHu}.)  We are able
to model the statistical entropy of such near-extremal black holes as
a boosted dilute gas of massless particles.

    Each of the two-form field strengths in a maximal supergravity can
be used to construct a single-charge black hole solution.  These
solutions form a multiplet under the Weyl subgroup of the U-duality
group \cite{lpsweyl}.  In $D=7$, 9 and 6, certain members of the
multiplet can be double-dimensionally oxidised to become the D3-brane,
M2-brane and M5-brane respectively.  The entropy of such a black hole
in the near-extremal regime can then be modelled by a dilute gas in
the world-volume of the corresponding $p$-brane.  In $6\le D\le 9$,
one can construct 2-charge black holes that are generating solutions
for the most general black holes.  The 2-charge solutions associated
with different field configurations also form multiplets under the
Weyl subgroup of the U-duality group.  Some configurations can be
viewed as intersections of $p$-branes in higher dimensions.  In this
paper, we focus on the cases which correspond to ``boosted'' D3-brane
and M-branes.  In other words, the second charge is carried by the
Kaluza-Klein vector.  In these cases, we show that the contribution to
the entropy of the system due to the Kaluza-Klein charge can be
understood as a simple consequence of the Lorentz contraction
resulting from the boost.  Thus the previously-known dilute gas model
for the D3-brane and M-branes can be used to understand
microscopically the two-charge black holes, except that the dilute gas
is now in a boosted frame, rather than in the rest frame.  In
particular, when the momentum of the system is held fixed as the 
boost becomes large,
it is associated with the Kaluza-Klein charge in the
supergravity picture.  This observation is consistent with the
conjecture that M-theory or type IIB theory on K$_D\times S^P$ is dual
to a CFT in an infinite-momentum frame.  It is worth remarking that
the approach we have adopted here for studying 2-charge black holes,
by considering the case where one of the charges is carried by a
Kaluza-Klein vector, seems to be the easiest way of tackling the
problem.  Other 2-charge black holes, whose higher-dimensional
interpretation would be as intersections of $p$-branes, are related to
the ones we study here by U-duality transformations.  The method we
adopt here, combined with U-duality, seems to provide the easiest way
for providing an interpretation for the entropy of the black holes
that correspond to intersections of $p$-branes,

     The paper is organised as follows.  In section 2, we discuss the
extremal and non-extremal M2-branes, with the inclusion of a pp-wave,
and show how their dimensional reductions on $S^7$ give rise to the
four-dimensional Kaigorodov and Carter-Novotn\'y-Horsk\'y metrics
respectively.  We compare the entropy and temperature of the $D=9$
two-charge black hole obtained by compactification on $T^2$ with the
corresponding results for the $S^7$ compactification, and show that
they agree in the near-extremal regime.  In section 3, we generalise
the results to the case of the M5-brane and the D3-brane.  In section
4, we give a brief discussion of reductions to $D=3$ and $D=2$, which
include in particular the BTZ black hole in $D=3$.  Appendix A
contains our results for the generalisation of the Kaigorodov metric
to arbitrary spacetime dimensions, and in Appendix B we construct its
Killing vectors and Killing spinors.  In Appendix C we generalise the
Carter-Novotn\'y-Horsk\'y metric to arbitrary dimensions, and
construct its Killing vectors.

\section{M2-brane with a pp-wave\label{sec:boostm2}}

\subsection{Extremal case\label{sec:bm2extr}}

       We first consider the intersection of an extremal M2-brane and a
gravitational pp-wave in $D=11$ supergravity.  The classical solution
is given by
\bea
ds_{11}^2 &=& H^{-2/3}(-K^{-1}\, dt^2 + K\, (dx_1 + (K^{-1}-1)\, dt)^2
+ dx_2^2)  + H^{1/3}\, (dr^2 + r^2\, d\Omega_7^2)\ ,\nn\\
F_4&=&dt\wedge dx_1 \wedge dx_2 \wedge dH^{-1}\ ,\label{m2wave}\\
H&=& 1 + \fft{Q_1}{r^6}\ ,\qquad
K= 1 + \fft{Q_2}{r^6}\ .\nn
\eea
It is worth mentioning that the harmonic function of a single wave in
$D=11$, which would give rise to a D0-brane in $D=10$, would depend on
$(r^2 + x_2^2)^{7/2}$ rather than just $r^6$.  The above solution
describes a pp-wave, uniformly distributed along the world-volume
coordinate $x_2$, and propagating in the direction of the world-volume
coordinate $x_1$.

     Performing a double-dimensional reduction on the spatial
coordinates $x_1$ and $x_2$, one obtains a 2-charge black hole in
$D=9$ maximal supergravity, with the two charges carried by the
winding vector $A_{\1 12}$, coming from the dimensional reduction of
$A_\3$ in $D=11$, and the Kaluza-Klein vector $\cA^1_\1$.  (In this
paper, we adopt the notation of \cite{lpsol,cjlp1} for the
lower-dimensional fields in maximal supergravities.)  Note that in
(\ref{m2wave}) we have, for simplicity, chosen the special case where
the wave propagates along the $x_1$ direction.  In general, the wave
can propagate in an arbitrary direction in the $(x_1,x_2)$ plane.
This is reflected in the fact that in $D=9$ maximal supergravity there
are two Kaluza-Klein vectors $\cA_\1^1$ and $\cA_\1^2$, which form a
doublet under the $GL(2,\R)$ global symmetry of $D=9$ maximal
supergravity.  To get the general solution, we can start with the
above simple 2-charge solution, involving $\{A_{\1 12}, \cA^1_\1 \}$,
and apply an $SL(2,\R)$ global symmetry transformation, under which
$A_{\1 12}$ is a singlet.  Then we oxidise the solution back to
$D=11$, and thus obtain the solution of the intersection of M2-brane
and a wave that propagates on a general direction in the world-volume
of the M2-brane.  However, since the $GL(2,\R)$ global symmetry is
nothing but the residual part of the internal general coordinate
transformations of $D=11$ supergravity, it follows that such a wave
propagating in a general world-volume direction can be obtained from
(\ref{m2wave}) by an appropriate general coordinate transformation.
It should be noted however that the general coordinate transformation
may have the effect of altering the global structure of the solution.

     We are interested in the near-horizon geometry of the M2-brane
with pp-wave (\ref{m2wave}).  The near horizon is defined to
be the regime where $Q_1/r^6 >>1$, and hence the membrane harmonic
function has the form $H\sim Q_1/r^6$ in this region.  Note that the
size of the non-vanishing wave charge (momentum) $Q_2$ is unimportant,
since we have $K\rightarrow K-1$ under the general coordinate
transformation \cite{hyun}
\be
t\longrightarrow \ft32 t -\ft12 x_1, \qquad
x_1 \longrightarrow \ft12 t + \ft12 x_1\ .\label{gencoord}
\ee
It is worth mentioning that this near-horizon structure can also be
obtained by a number of somewhat different procedures, using U-duality
symmetries or T-duality transformations to change the values of the
constant terms in the harmonic functions in any $p$-brane solution
\cite{hyun,bps,bb,cllpst}. (For the harmonic function $K$, as we have
seen, it can be achieved by a mere coordinate transformation.)  The
simplest way to remove the constant ``1'' in the harmonic function $H$
in (\ref{m2wave}) is to perform a dimensional reduction of
(\ref{m2wave}) on the entire set of three world-volume coordinates of
the M2-brane, including the time direction.  This gives rise to an
instanton solution of an eight-dimensional Euclidean-signatured
supergravity, which has an $SL(2,\R)$ symmetry that can be used to
rescale and shift the harmonic function $H$ by constants while leaving
the structure of the solution unaltered \cite{cllpst}.  Having
performed the symmetry transformation that leads to $H\rightarrow
H-1$, we can oxidise the solution back to $D=11$, obtaining the
near-horizon structure of (\ref{m2wave}).  It is not clear however
about the significance and the physical interpretation of such a
transformation.

     The metric of the near horizon of (\ref{m2wave}) is given by
\be
ds_{11}^2=Q_1^{-2/3} r^4\, (-K^{-1}\, dt^2 + 
K\, (dx_1 + (K^{-1} -1) dt)^2 + dx^2) + Q_2^{1/3} r^{-2} dr^2 +
Q_1^{1/3} d\Omega_7^2 \ .
\ee
Thus we see that the spacetime is a product $M_4\times S^7$.  It is of
interest to study this new vacuum of M-theory in more detail, and in
particular to study the structure of $M_4$.  Since the coefficient of
the $S^7$ metric $d\Omega_7^2$ is a constant, it follows that $M_4$
must be an Einstein metric, a solution of $D=4$ gravity with a pure
cosmological term:
\be
e^{-1}{\cal L}_4 = R- 2 \Lambda\ ,
\ee
with $\Lambda= -12 Q_1^{-1/3}$.  Here, we choose to take the
internal metric to be $ds_7^2 = Q_1^{1/3} d\Omega_7^2$.  After the
$S^7$ reduction, we obtain the four-dimensional metric
\bea
ds_4^2 = Q_1^{1/3} \Big(-e^{10\rho}\, dt^2 + e^{-2\rho}\, 
(dx_1 + e^{6\rho}\, dt)^2 + e^{4\rho}\, dx_2^2 + d\rho^2\Big)\ ,
\label{m4metric}
\eea
where $e^\rho=r$.  (Note that inside the parentheses we have absorbed
the charge parameters by rescaling the world-volume coordinates.  See 
\cite{bdlps} for a detailed discussion of spherical dimensional reduction.)

     It is straightforward to verify that (\ref{m4metric}) is an
homogeneous Einstein metric, but that it is not AdS$_4$; in fact, it
is a metric discovered first by Kaigorodov \cite{kaig}.  We shall
denote this metric as K$_4$.  In Appendices A and B, we discuss the
properties of this metric, and we derive its higher-dimensional
generalisations. Included in this is a discussion of the symmetries of
the generalised Kaigorodov metrics, and a construction of their
Killing spinors and Killing vectors.  To be specific, the K$_4$ metric
(\ref{m4metric}) has a 5-dimensional isometry group, and it preserves
$1/4$ of the supersymmetry.

\subsection{Decoupling limit}

         It is instructive to study whether there also exists a limit
in this M2-brane/wave solution where the field theory on the brane
decouples from the bulk.  To do this, we note that the metric in
(\ref{m2wave}) can be expressed, after the coordinate transformation
(\ref{gencoord}), as
\be
ds_{11}^2 = H^{-2/3}\, (K\, dx_1^2 + 2 dx_1\, dt + dx_2^2) +
     H^{1/3}\, (dr^2 + r^2 d\Omega_7^2)\ ,\label{m2wave2}
\ee
where $H=1 + Q_1/r^6$ and $K=Q_2/r^6$.  The membrane charge $Q_1$ is
subject to the Dirac quantisation condition in the presence of a
5-brane.  This implies that $Q_1= N\, \ell_p^6$, where $N$ is an
integer and we define the eleven-dimensional Plank length $\ell_p=
\kappa_{11}^{2/9}$.  The charge $Q_1$ is associated with a momentum
density $P$, {\it viz}, $Q_1\sim P\,\ell_p^9$.  In the asymptotic
region $r\rightarrow \infty$, the solution (\ref{m2wave2}) is
Minkowskian, {\it i.e.}\ $ds^2= 2dx_1\, dt + dx_2^2 + dr^2 + r^2
d\Omega_7^2$.  Note that in this region $x_1$ and $t$ become
light-cone coordinates.

    Following \cite{mald}, we consider the limit $\ell_p \rightarrow
0$, while keeping $U=2r^2/(N\, \ell_p^3)$ fixed.  In this limit,
we have
\be
N\ell_p^6/r^6 >>1\ ,\label{m2limit1}
\ee
and hence we can ignore the constant 1 in $H$. The
metric (\ref{m2wave2}) becomes
\be
ds_{11}^2 = \ell_p^2\, N^{1/3}\, \Big(
\ft14 \Big[\fft{8P}{N^{3/2}}\, \fft{dx_1^2}{U} +
U^2\, (2dx_1\, dt + dx_2^2) +
\fft{dU^2}{U^2}\Big] + d\Omega_7^2\Big)\ .\label{m2wave3}
\ee
Note that the radius of the Kaigorodov metric is half of that of the
seven-sphere. If the momentum density $P$ of the wave vanishes, then
$ds^2/\ell_p^2$ is a metric on AdS$_4\times S^7$ that depends only on
$N$, but is independent of $\ell_p$.  The limit where gravity
decouples is achieved by taking $\ell_p$ to approach zero \cite{mald}.
In our case, in order instead to maintain the form of the Kaigorodov
metric, the momentum density $P$ must be non-vanishing and fixed.  The
metric $ds^2/\ell_p^2$ then becomes K$_4\times S^7$, which is
independent on $\ell_p$.

      Thus we see that the decoupling limits in the two cases of the
M2-brane and the boosted M2-brane are the same, and in both cases the
radius of the seven-sphere is the same, namely $R_7=N^{1/3}\, \ell_p$.
Furthermore, in both cases the momentum density is fixed, but with the
difference that in the AdS case the momentum density is zero, whilst
in the Kaigorodov case the momentum is non-vanishing.  It was
conjectured in \cite{mald} that M-theory on AdS$_4\times S^7$ is dual
to a $2+1$ dimensional conformal theory.  In the case of K$_4\times
S^7$, the K$_4$ can be viewed as infinitely-boosted AdS$_4$, and the
gravitational decoupling limit that maintains the Kaigorodov metric
requires that the momentum density remains fixed and non-vanishing.
We expect that M-theory on such a metric is dual to the conformal
field theory in the infinitely-boosted frame, with constant momentum
density.  A natural consequence of this conjecture is that the entropy
of the boosted M2-brane in the near-extremal regime can be modelled by
a dilute gas in a highly-boosted frame for the three-dimensional
spacetime, with constant momentum density.  We shall show later that
this is indeed the case.

\subsection{Non-extremal case}

     We now turn our attention to the non-extremal M2-brane with a
superimposed gravitational pp-wave.  The solution is given by
\bea       
ds_{11}^2 &=& H^{-2/3}(-K^{-1}\, e^{2f}\, dt^2 +
    K(dx_1 + \coth\mu_2\, (K^{-1}-1)dt)^2 + dx_2^2) \nn\\
 &&+ H^{1/3}(e^{-2f} dr^2 + r^2 d\Omega_7^2)\ ,\nn\\
A_\3 &=& \coth\mu_1\, H^{-1}\, dt\wedge dx_1 
\wedge dx_2\ ,\label{nonextm2}
\eea
where 
\bea
H=1 + \fft{\kappa_{11}^{4/3}\, k}{r^6}\, \sinh^2\mu_1\ ,\qquad
K=1 + \fft{\kappa_{11}^{4/3}\, k}{r^6}\, \sinh^2\mu_2\ ,\qquad
e^{2f} = 1 -\fft{\kappa_{11}^{4/3}\,k}{r^6}\ ,\label{d11hkf}
\eea
The horizon of the boosted M2-brane is at $r_+=\kappa_{11}^{2/9}\,
k^{1/6}$.  Note that in this non-extremal case, the effect of the
superimposed pp-wave can be removed by a coordinate transformation.
Specifically, the coordinate transformation (\ref{boost}) given in
Appendix C maps the metric (\ref{nonextm2}) into the unboosted
non-extremal M2-brane, with the metric
\be       
ds_{11}^2 = H^{-2/3}(-e^{2f}\, d{t'}^2 + d{x_1'}^2 + dx_2^2)
 +  H^{1/3}(e^{-2f} dr^2 + r^2 d\Omega_7^2)\ .\label{unboostedm2}
\ee
Note that the transformation (\ref{boost}) is incompatible with any
periodic identification of the $x_1$ coordinate.  Therefore it is only
in the case of a wave propagating along an infinite (\ie unwrapped)
world-volume direction on the M2-brane that it can be transformed into
a solution with no wave.  Note also that the coordinate transformation
(\ref{boost}), which corresponds to a Lorentz boost in the $(t,x_1)$
plane with velocity $\tanh\mu_2$ (see (\ref{lorboost})), becomes
singular in the extremal limit where $\mu_2\rightarrow\infty$.

   The Hawking temperature and entropy per unit 2-volume of the metric
(\ref{nonextm2}) are easily calculated to be
\bea
T&=&\fft{3}{2\pi r_+}\, (\cosh\mu_1\, \cosh\mu_2)^{-1}\ ,\nn\\
\fft{S}{L_1\,L_2}&=&\fft{\hbox{Area}}{4\kappa_{11}^2\, L_1\, L_2}\ \nn\\
 &=& \fft{k^{7/6}}{4\k_{11}^{4/9}}\, 
\Omega_7\, \cosh\mu_1\, \cosh\mu_2\ .\label{d11entropy}
\eea
The physical interpretations of the $\mu_1$ and $\mu_2$ dependences in
the entropy formula in (\ref{d11entropy}) are quite different.  When
there is no boost, {\it i.e.}\ $\mu_2=0$, in the near-extremal regime,
the entropy can be modelled as dilute gas in the M2-brane world-volume
\cite{kt,kt1}, as we shall review presently.  On the other hand, the
$\mu_2$ dependence is more easily understood.  In fact, as we discuss
in appendix C, $\cosh\mu_2$ is precisely the $\gamma$-factor of the
Lorentz boost (\ref{boost}) along the $x_1$ direction, associated with
the propagation of the pp-wave (see (\ref{lorboost})).  The entropy of
a closed system, which is a measure of the distribution of occupancy
numbers of the states, is a Lorentz invariant quantity.  However, the
entropy density $S/(L_1\, L_2)$ by contrast is not Lorentz invariant,
since under the Lorentz boost we have $L_1 \rightarrow
L_1'=\gamma^{-1}\, L_1$ and $L_2 \rightarrow L_2'=L_2$.  It follows
that under the boost, the new entropy density becomes $S/(L_1'\,
L_2')$, which is $\gamma$ times the original density.  Thus after a
notational change, in which the primed periods $L_i'$ are replaced by
the unprimed periods $L_i$, we obtain the entropy formula given in
(\ref{d11entropy}).  This provides a simple explanation for how the
entropy depends on the extra charge.\footnote{Note also that the fact
that the charge $Q_2$ has an interpretation as the momentum {\it
density} of the wave in the higher dimension is also easily
understood.  Upon performing the Lorentz boost, the transformation
from the zero-momentum frame to a frame with velocity $v=\tanh\mu_2$
gives a momentum proportional to $\gamma\,v = \cosh\mu_2\, \tanh\mu_2
= \sinh\mu_2$, and the momentum density acquires a further
$\cosh\mu_2$ dilatation factor, giving an overall $\sinh2\mu_2$
dependence.  This is exactly the way in which the charge depends on
$\mu_2$, as can be seen from (\ref{mQs}).}  In particular, in the
near-extremal regime the microscopic entropy of the boosted M2-brane
can be modelled by a dilute massless gas in a boosted frame with boost
parameter $\gamma=\cosh\mu_2$.

\subsection{$T^2$ reduction}

    Since the solution has translational isometries on the
world-volume spatial coordinates $(x_1, x_2)$, we can perform
dimensional reductions on these two coordinates, thereby obtaining a
2-charge non-extremal isotropic black hole in $D=9$.  The relevant
part of the $D=9$ dimensional Lagrangian that describes this solution
is \cite{lpsol,cjlp1}
\be
{e^{-1} \kappa_9^2}\, {\cal L} 
= R -\ft12(\del\vec\phi)^2 - \ft14 e^{\vec a_{12} \cdot
\vec \phi}\, (F_{\2 12})^2 -\ft14 e^{\vec b_1 \cdot \vec \phi}\,
(\cF_\2^1)^2\ ,
\ee
where $\vec \phi=(\phi_1, \phi_2)$, $\vec a_{12} = (1, 3/\sqrt7)$ and
$\vec b_1 = (-3/2, -1/(2\sqrt7))$.  The $D=9$ dimensional gravitational
constant is related to the one in $D=11$ as follows
\bea
\kappa_9^2 = \fft{\k_{11}^2}{L_1 L_2}\ ,
\eea
where $L_1$ and $L_2$ are the periods of the coordinates $x_1$ and
$x_2$ respectively.  The non-extremal $D=9$ black hole solution is
\cite{dlpblack,ct}
\bea
&&ds_9^2 = -(H K)^{-6/7}\, e^{2f}\, dt^2 + (HK)^{1/7}(e^{-2f}\, dr^2 +
r^2 d\Omega_7^2)\ ,\nn\\
&&\vec \phi = \ft12 \vec a_{12}\, \log H + \ft12 \vec b_1\, 
\log K\ ,\nn\\
&&A_{\1 12} = \coth\mu_1\, H^{-1}\, dt, 
\qquad \cA_{\1}^1 = \coth\mu_2\, K^{-1}\, dt\ ,
\label{d9bh}
\eea
which has  mass and charges given by
\bea
M= k\, \kappa_{11}^{4/3}(6 \sinh^2\mu_1 + 6 \sinh^2\mu_2 + 7)
L_1\, L_2\, \Omega_7 \ ,\nn\\
Q_1 = 3k\, \kappa_{11}^{4/3}\, 
 \sinh2\mu_1\ ,\qquad Q_2 = 3k\,\kappa_{11}^{4/3} \, \sinh2\mu_2\ ,
\label{mQs}
\eea
where $\Omega_7$ is the volume of the unit seven-sphere.

   The Hawking temperature and entropy are easily calculated to be
\bea
T&=&\fft{3}{2\pi r_+}\, (\cosh\mu_1\, \cosh\mu_2)^{-1}\ ,\nn\\
S&=&\fft1{4\kappa_9^2}\, r_+^7\, 
\Omega_7\, \cosh\mu_1\, \cosh\mu_2
 = \fft{k^{7/6}\, L_1\, L_2}{4\k_{11}^{4/9}}\, 
\Omega_7\, \cosh\mu_1\, \cosh\mu_2\ .\label{d9entropy}
\eea
Note that the entropy and temperature are the same as those given in
(\ref{d11entropy}).

       The number of charges of generating solutions of the most general
black holes in $D=5$ and $D=4$ are $N=3$ and $N=4$
respectively;\footnote{Actually, in $D=4$ the generating solution for
the {\sl most} general black holes is specified by $N=5$ charges;
however, the metric is still of the Reissner-Nordstr{\oo}m-type
\cite{CvTsI}.} their global spacetime structure is the same as that of
the Reissner-Nordstr{\oo}m black holes, since the moduli of these
solutions are finite near the horizon. As a consequence, the entropy
is non-vanishing even in the extremal limit.  In $D\ge 6$, the number
of charges of generating black hole solutions is $N=2$. These black
holes are dilatonic, meaning that the dilaton diverges on the horizon
in the extremal limit, and consequently the entropy vanishes in the
extremal limit.  Thus the entropy in the near-extremal regime can best
be characterised by its relation to the temperature. The single-charge
black holes in $D=9$, $D=7$ and $D=6$ can be viewed as being
essentially equivalent to the M2-brane, D3-brane and M5-brane, since
they are the double-dimensional reductions of these branes.  The
entropy and temperature satisfy the ideal-gas relationship $S\sim T^p$
in the near extremal regime, where $p$ is the world-volume spatial
dimension of the M-brane or D3-brane \cite{kt,kt1}.\footnote{The
relation is obtained by using the expressions given in
(\ref{d9entropy}) for the entropy and temperature of the black hole,
with $\mu_2$ set to zero (so that there is no boost charge) and taking
the limit where $\mu_1$ becomes large, while keeping the charge $Q_1$,
given in (\ref{mQs}), fixed.  The approximation becomes a good one
when $\cosh\mu_1$ and $\sinh\mu_1$ can be approximated by $\ft12
e^{\mu_1}$.}

        When the M2-brane contains also a pp-wave, and hence gives
rise to a 2-charge black hole in $D=9$, the relation changes
drastically; it becomes instead $S\sim T^{1/5}$.  (The calculation in
this case is similar to the previous one, except that now $\mu_2$ is
also taken to be large, and both the charges $Q_1$ and $Q_2$ are held
fixed.)  In general, a 2-charge black hole in maximal supergravity has
entropy and temperature that satisfy the relation $S\sim T^{1/(D-4)}$.
As we discussed below (\ref{d11entropy}), the entropy density formula
with the inclusion of the extra Kaluza-Klein charge is precisely
accounted for by the effect of the Lorentz contraction along the
direction of the boost.  

         In the next subsection, we shall show that the calculation of
the entropy of the 2-charge $D=9$ black hole in the near-extremal
regime can be mapped into a calculation of the entropy of the
Carter-Novotn\'y-Horsk\'y solution to $D=4$ gravity with a pure
cosmological constant.

\subsection{$S^7$ reduction}

        We may now look at the M2-brane with pp-wave from another
angle.  Note that near to the horizon, we have $H=1+ (r_+/r)^6\,
\sinh^2\mu_1$.  It follows that we have $H\sim (r_+/r)^6\,
\sinh^2\mu_1$ in this region, provided that the solution is nearly
extremal, namely $\mu_1 >>1$.  In other words, in the near-extremal
regime the 1 in harmonic function $H$ can be dropped near the horizon.
(As in the case of extremal solutions, the 1 in the harmonic function
can in fact be removed by U-duality or T-duality transformations.)  As
in the extremal case, the metric then becomes a product, $M_4\times
S^7$.  Thus we can compactify on the 7-sphere, and obtain a
configuration in $D=4$ that is a solution of Einstein gravity with a
pure cosmological term.

         It is convenient to take the internal $S^7$ metric to be
$ds_7^2 = \k_{11}^{4/9}\, (k\sinh^2\mu_1)^{1/3}\, d\Omega_7^2$, {\it
i.e.}\ the radius of the $S^7$ is 
\be
R_7=\k_{11}^{2/9}\, (k\sinh^2\mu_1)^{1/6}\ .
\ee
It follows that the four-dimensional gravitational constant is given
by
\be
\k_4^2 = \fft{\k_{11}^2}{V_{S^7}} =
\fft{\k_{11}^{4/9}}{(k\sinh^2\mu_1)^{7/6}\, \Omega_7}\ .
\ee
The four-dimensional metric resulting from the $S^7$ reduction is given by
\bea
ds_4^2 &=& (\kappa_{11}^{4/3}\, k\, \sinh^2\mu_1)^{-2/3}\,  r^4\, 
(-K^{-1}\, e^{2f}\, dt^2 + K(dx_1 + \coth\mu_2\, 
(K^{-1} -1)\, dt)^2 + dx_2^2)\nn\\ 
&&+ (\kappa_{11}^{4/3}\, k\, \sinh^2\mu_1)^{1/3}\, e^{-2f}\, 
\fft{dr^2}{r^2} \ .\label{d4nonext}
\eea
This is a solution to the Einstein equations coming from the
Lagrangian $e^{-1}\, \k^4\, {\cal L} = R - 2 \Lambda$, where $\Lambda=
-12(\kappa_{11}^{4/3}\, k\, \sinh^2\mu_1)^{-1/3}$.  The metric
(\ref{d4nonext}) is no longer homogeneous when the non-extremality
factor $e^{2f}$ is present.  It is in fact an Einstein metric found by
Carter \cite{carter}, and by Novotn\'y and Horsk\'y
\cite{novhor}. (See also \cite{grbook}).)  In the asymptotic regime
$r\rightarrow \infty$ we have $e^{2f}\rightarrow 1$, and the metric
becomes the Kaigorodov metric, discussed in the previous section.

      The entropy of the solution (\ref{d4nonext}) is given by a
quarter of the area of the horizon, which is spanned by the spatial
coordinates $x_1$ and $x_2$:
\bea
S&=&\fft{\hbox{Area}}{4\k_4^2}\nn\\
 &=&\fft{L_1 L_2}{4\k_{11}^{4/9}}\, k^{7/6}\, 
\Omega_7\, \sinh\mu_1\, \cosh\mu_2\ .\label{d4entropy}
\eea
In the near-extremal limit $\mu_1 >>1$ we have $\sinh\mu_1 \sim
\cosh\mu_1$, and hence the {\it four}-dimensional entropy
(\ref{d4entropy}) is the same as the entropy (\ref{d9entropy}) of the
{\it nine}-dimensional black hole.  Analogously, the Hawking
temperature that one calculates for the four-dimensional metric
(\ref{d4nonext}) is the same as the expression given in
(\ref{d9entropy}) for the $D=9$ black hole, except that again the
$\cosh\mu_1$ factor is replaced instead by $\sinh\mu_1$.  Again, this
means that the $D=9$ black-hole calculation and the $D=4$ calculation
with the Carter-Novotn\'y-Horsk\'y metric agree in the near-extremal
limit when $\mu_1>>1$.

       The agreement of the entropies stems from the fact that
dimensional reduction and oxidation leave entropies invariant.  To see
this, let us consider a metric in $\wtd D$ dimensions with an horizon
of area $A_{\wtd D}$.  The entropy is then given by $S=\wtd A_{\wtd
D}/(4\k_{\wtd D}^2)$.  If we perform a dimensional reduction on an
internal space that has volume $V$, to give rise to a $D$-dimensional
metric, then the area of the horizon is $A=\wtd A/V$, and hence the
entropy becomes $S=A/(4\k_D^2)=\wtd A/(4\k_D^2\, V)$.  Thus the
entropy is preserved under dimensional reduction, since $\k_{\wtd D}^2
= \k_{D}^2\, V$.  It is for this same reason that the entropy of black
holes in $D=5$ and $D=4$ can be mapped to the problem of BTZ black
holes in $D=3$.  Of course the agreement that we are seeing here
operates only in the near-extremal regime where $\mu_1>>1$, since we
made the approximation that the constant term in the harmonic function
$H$ could be neglected, in our derivation of the four-dimensional
Carter-Novotn\'y-Horsk\'y metric by reduction on the
seven-sphere.\footnote{In \cite{hyun,ss}, the solutions for
Reissner-Nordstr{\oo}m-type black holes in $D=5$ and $D=4$ were mapped
into three-dimensional BTZ solutions, where the entropy was shown to
agree with the original black-hole results in $D=5$ and $D=4$.  The
mappings were implemented using duality symmetries to shift the
constant term in the harmonic function $H$ to zero.  Although it was
shown in \cite{hyun,ss} that a mapping could be found that leads to an {\it
exact} agreement for the two entropy calculations, it would seem that
other valid mappings could instead have been performed for which the
agreement would be seen only in the near-extremal limit.}

\subsection{Microscopic entropy and boosted dilute gas}

       The correspondence of M-theory on K$_4\times S^7$ and CFT on an
infinite momentum frame provides a possible microscopic interpretation
of 2-charge black holes in $D=9$.  As we have mentioned, in the case
where there is no pp-wave propagating on the M2-brane, the entropy and
temperature in the near-extremal regime satisfy the ideal-gas relation
$S\sim T^2$ in two-dimensional space \cite{kt}.  When the pp-wave is
superimposed in the M2-brane, the entropy is a natural consequence of
the Lorentz contraction along the direction of the associated boost,
which leads to the appropriate dilation of the entropy density, as
discussed in detail in section 2.3. Thus in the near-extremal regime,
the entropy can be modelled microscopically as a dilute massless gas
in a boosted frame, {\it viz.}  $S=\cosh\mu_2\, S_{\rm dilute\,\,
gas}$.  This formula applies to any boost.  If the boost is finite,
then in the limit towards extremality, the momentum of the system
becomes zero, and it corresponds to the single-charge solution.  If on
the other hand, we hold the momentum density $k\, e^{2\mu_2}$ fixed
and finite while boosting the system towards speed of light, it
corresponds to the near-extremal regime of the 2-charge black hole in
$D=9$, and the extra charge is associated with the momentum.  Note
that this is a natural consequence of the conjecture that M-theory on
the K$_4\times S^7$ background is dual to the $2+1$ dimensional
conformal field theory in an infinitely-boosted frame with constant
momentum density, where K$_4$ is the four-dimensional Kaigorodov
metric.

\section{M5-brane or D3-brane with a 
pp-wave\label{sec:m5d3}}

\subsection{M5-brane\label{sec:m5}}

     The discussion in the previous section can be equally applied to
the M5-brane and the D3-brane.  We shall first look at the extremal
M5-brane in the presence of a pp-wave.  The supergravity solution is
given by
\bea
ds_{11}^2 &=& H^{-1/3}(-K^{-1}\, dt^2 + K\, (dx_1 + (K^{-1}-1)\, dt)^2
+ dx_2^2+ \cdots + dx_5^2)\nn\\
&&  + H^{2/3}\, (dr^2 + r^2\, d\Omega_4^2)\ ,\nn\\
F_4&=& {*(dH^{-1}\wedge d^6x)},\label{m5wave}\\
H&=& 1 + \fft{Q_1}{r^3}\ ,\qquad
K= 1 + \fft{Q_2}{r^3}\ ,\nn
\eea
where $d^6x$ is the volume form on the 5-brane world-volume.
To be precise, the solution describes a 5-brane in $D=11$, with a
wave, uniformly distributed on the world-volume coordinates $(x_2,
\ldots, x_5)$, and propagating in the world-volume direction $x_1$.

       The dimensional reduction of (\ref{m5wave}) on all five spatial
5-brane world-volume coordinates gives rise to a 2-charge black hole
in $D=6$.  One charge is the magnetic charge carried by the 4-form
field strength $F_4$, and the other is the electric charge carried by
the Kaluza-Klein 2-form field strength $\cF_\2^1$.  We are also
interested in the $S^4$ reduction, and, in particular, the reduction
in the near-horizon limit $r\rightarrow 0$.  In this regime, we have
$Q_1/r^3>>1$, and hence the constant 1 in the harmonic function $H$
can be dropped.  The space then becomes a product K$_7\times S^4$,
{\it viz}
\bea
ds_{11}^2&=&Q_1^{-1/3} r\, (-K^{-1}\, dt^2 + 
K\, (dx_1 + (K^{-1} -1) dt)^2 + dx_2^2 + \cdots + dx_5^2)\nn\\
&& + Q_1^{2/3} r^{-2} dr^2 +Q_1^{2/3} d\Omega_4^2\ .
\eea
Compactifying the solution on the $S^4$, with $ds_4^2 = Q_1^{2/3}\,
d\Omega_4^2$, we obtain the seven-dimensional Einstein metric 
\be
ds_7^2 = Q_1^{2/3}\, \Big( -e^{4\rho}\, dt^2 + e^{-2\rho}\, (dx_1+
e^{3\rho}\, dt)^2 + e^{\rho}\, (dx_2^2 + \cdots dx_5^2) +
d\rho^2\Big)\ .\label{d7kaig} 
\ee 
This is precisely the generalised Kaigorodov metric in $D=7$, which is
derived and its properties discussed in detail in Appendices A and B.
The metric (\ref{d7kaig}), which we denote by K$_7$, is homogeneous
and Einstein, and is a solution to $D=7$ gravity with a pure
cosmological term\footnote{Note that in $D$ dimensions, the
Einstein-Hilbert action with cosmological term ${\cal L} = e\, R - e\,
(D-2)\, \Lambda$ gives rise to the Einstein equations $R_{\mu\nu} =
\Lambda\, g_{\mu\nu}$.  Thus the somewhat unusual-looking
normalisation for the cosmological term in the action is needed in
order to have a canonical-looking form for the Ricci tensor for the
Einstein metric.} $e^{-1} {\cal L}_7 =R-5\Lambda$ with $\Lambda= -24\,
Q_1^{-2/3}$.

     We shall now consider the limit where the dynamics of the 5-brane
decouples from the bulk.  Note that we have $Q_1=N \ell_p^3$ and $Q_2
= P\,\ell_p^9$, where $P$ is the momentum density of the
five-dimensional world spatial volume.  Following \cite{mald}, we
consider the limit $\ell_p\rightarrow 0$ with
$U=\ft12\sqrt{r/(N\,\ell_p^3)}$ fixed.  In this case, we have that
$N\ell_p^3/r^3 >>1$, and hence 1 in function H can be dropped, giving
rise to the metric
\be
ds_{11}^2 =\ell_p^2\, N^{2/3} \Big( 4\Big[\fft{P}{64N^3}\, 
\fft{dx_1^2}{U^4} +
U^2\,(2dx_1\, dt + dx_2^2 + \cdots + dx_5^2) +
\fft{dU^2}{U^2}\Big] + d\Omega_4^2\Big)\ .
\ee
Thus the decoupling limit for the M5-brane/wave system is the same as
for the pure M5-brane, but giving rise to K$_7\times S^4$ or
AdS$_7\times S^4$ respectively.  In the latter case, the momentum
density $P$ vanishes, whilst in the former case the momentum density
is fixed but non-vanishing. Thus we expect that M-theory on K$_7\times
S^4$ is dual to the $(0,2)$ conformal theory in an infinitely-boosted
frame, with constant momentum density.

        The discussion for the non-extremal case is straightforward.
The solution in $D=11$ is given by
\bea       
ds_{11}^2 &=& H^{-1/3}(-K^{-1}\, e^{2f}\, dt^2 +
    K(dx_1 + \coth\mu_2\, (K^{-1}-1)dt)^2 + dx_2^2 + \cdots +
dx_5^2) \nn\\
 && + H^{2/3}(e^{-2f} dr^2 + r^2 d\Omega_4^2)\ ,\nn\\
F_\4 &=& \coth\mu_1\, {*(dH^{-1}\wedge d^6x)}\ ,
\eea
where 
\bea
H=1 + \fft{\kappa_{11}^{2/3}\, k}{r^3}\, \sinh^2\mu_1\ ,\qquad
K=1 + \fft{\kappa_{11}^{2/3}\, k}{r^3}\, \sinh^2\mu_2\ ,\qquad
e^{2f} = 1 -\fft{\kappa_{11}^{2/3}\,k}{r^3}\ ,\label{d11hkf2}
\eea
The horizon of the boosted M5-brane is at $r_+=\kappa_{11}^{2/9}\,
k^{1/3}$.  (Again, as we discussed for the M2-brane, locally in this
non-extremal case the harmonic function $K$ associated with the wave
can be set to 1 by the coordinate transformation (\ref{boost}).)

      First, let us consider the double-dimensional reduction on the
world-volume coordinates $(x_1, x_2, \ldots, x_5)$.  This gives rise to
a 2-charge non-extremal isotropic black hole in $D=6$.  The relevant
Lagrangian is given by \cite{lpsol,cjlp1}
\be
e^{-1}\,\kappa_6^2\, {\cal L}_6 =
R-\ft12(\del \vec \phi)^2 -\ft1{48} e^{\vec a\cdot \vec\phi}\, 
(F_\4)^2 -\ft14 e^{\vec b_1 \cdot \vec \phi}\, (\cF_\2^1)^2\ ,
\ee
where $\vec\phi=(\phi_1, \ldots, \phi_5)$ and
\bea \vec a&=&(-\ft12, -3/(2\sqrt7), -\sqrt{3/7}, -\sqrt{3/5})\ ,\nn\\
\vec b_1 &=& (-\ft32, -1/(2\sqrt7), -1/\sqrt{21}, -1/\sqrt{15})\ .
\eea
The six-dimensional gravitational constant is given by $\kappa_6^2 =
\kappa_{11}^2/(L_1\cdots L_5)$, where $L_i$ is the period of the
coordinate $x_i$.

       The six-dimensional non-extremal black hole is given by
\cite{dlpblack,ct}
\bea
&&ds_{6}^2=-(H\,K)^{3/4}\, e^{2f}\, dt^2 +
(H\,K)^{1/4}(e^{-2f}\, dr^2 + r^2 d\Omega_4^2)\ ,\nn\\
&& \vec \phi -\ft12 \vec a\, \log H + 
\ft12 \vec b_1\, \log K\ ,\nn\\
&& F_\4 =\coth\mu_1\,  e^{-\vec a\cdot\vec\phi}\, {*(dH^{-1}\wedge dt)}
\qquad {\cal A}_\1^2 = \coth\mu_2\, K^{-1}\, dt\ .
\label{d6bh}
\eea
It is straightforward to see that its entropy is
\bea
S&=& \fft{\hbox{Area}}{4\kappa_6^2}\ ,\nn\\
&=& \fft{L_1\cdots L_5}{\kappa_{11}^{10/9}}\, \Omega_4\, k^{4/3}\,
\cosh\mu_1 \, \cosh\mu_2\ .\label{d6ent}
\eea

       As in the extremal case, we can also consider the $S^4$
reduction of the non-extremal boosted M5-brane.  In particular, we
consider the near-extremal case, for which the constant 1 in the
harmonic function $H$ can be neglected near the horizon.  The space
becomes a product $M_7\times S^4$.  The metric of the internal
4-sphere is $ds_4^2 = \kappa_{11}^{4/9}\, (k\sinh^2\mu_1)^{2/3}\,
d\Omega_4^2$, and
thus its radius is $R_4= \kappa_{11}^{2/9}\, (k\sinh^2\mu_1)^{1/3}$.
It follows that the seven-dimensional gravitational constant is given by
\be
\kappa_7^2 = \fft{\kappa_{11}^2}{V_{S^4}} = 
\fft{\kappa_{11}^{10/9}}{(k\sinh^2\mu_1)^{4/3}\, \Omega_4}\ .
\ee
 
     By performing a dimensional reduction on the four-sphere we arrive at 
the seven-dimensional metric
\bea
ds_7^2 &=& (\kappa_{11}^{2/3}\, k\, \sinh^2\mu_1)^{-1/3}\,  r\, 
\Big(-K^{-1}\, e^{2f}\, dt^2 + K(dx_1 + \coth\mu_2\, 
(K^{-1} -1)\, dt)^2\nn\\
&& + dx_2^2 + \cdots + dx_5^2\Big)  + 
(\kappa_{11}^{2/3}\, k\, \sinh^2\mu_1)^{2/3}\, e^{-2f}\, 
\fft{dr^2}{r^2} \ .\label{d7nonext}
\eea
This solution is still an Einstein metric, but it is no longer
homogeneous.  It is in fact the seven-dimensional generalisation of
the Carter-Novotn\'y-Horsk\'y metric, which we obtain in Appendix C.
In the asymptotic regime $r\rightarrow \infty$ the metric approaches
the generalised homogeneous Kaigorodov metric discussed in Appendices
A and B.  The entropy of the metric (\ref{d7nonext}) is given by
\bea
S &=& \fft{\hbox{Area}}{4\kappa_7^2}\nn\\
  &=& \fft{L_1\cdots L_5}{\kappa_{11}^{10/9}}\, \Omega_4\, k^{4/3}\,
\sinh\mu_1 \, \cosh\mu_2\ ,\label{d7dent}
\eea
which agrees with (\ref{d6ent}) in the near-extremal limit $\mu_1
>>1$.  Note that when there is no boost on the M5-brane, the entropy
and temperature satisfy the ideal-gas relation $S\sim T^5$ of
five-dimensional space, in the near-extremal regime \cite{kt}.  When
the solution is largely-boosted, this relation becomes $S\sim
T^{1/2}$.  The $\mu_2$ dependence of the entropy density is again the
natural consequence of the Lorentz contraction on the world-volume,
associated with the boost, and hence the entropy density is enlarged
by $\cosh\mu_2$, which is the $\gamma$-factor of the Lorentz boost.
Thus the near-extremal entropy can be modelled by a dilute massless
gas in a boosted frame.  A particular interesting case is to highly
boost the dilute gas while hold the momentum density $k\, e^{2\mu_2}$
fixed.  This corresponds to the near-extremal 2-charge black holes in
$D=6$.  This boosted dilute gas model of the 2-charge black hole
entropy is consistent with the conjecture that M-theory on K$_7\times
S^4$ is dual to the CFT in an infinitely-boosted frame, with constant
momentum density.

\subsection{D3-brane\label{sec:d3}}

      The D3-brane is supported by the self-dual 5-form in the type IIB
theory.  The solution for an extremal D3-brane in the presence of a
gravitational pp-wave is given by
\bea
ds_{11}^2 &=& H^{-1/2}(-K^{-1}\, dt^2 + K\, (dx_1 + (K^{-1}-1)\, dt)^2
+ dx_2^2+ dx_3^2)\nn\\
&&  + H^{1/2}\, (dr^2 + r^2\, d\Omega_5^2)\ ,\nn\\
F_5&=& dH^{-1}\wedge d^4x + {*(dH^{-1}\wedge d^4x)},\label{d3wave}\\
H&=& 1 + \fft{Q_1}{r^4}\ ,\qquad
K= 1 + \fft{Q_2}{r^4}\ ,\nn
\eea
where $d^4x$ is the volume form on the world-volume of the D3-brane.
Note that the wave is uniformly distributed on the plane $(x_2, x_3)$
in the world-volume, and it propagates along the $x_1$ direction.

       The dimensional reduction of (\ref{d3wave}) on all three
spatial 3-brane world-volume coordinates gives rise to a 2-charge
black hole in $D=7$.  In the case instead of the $S^5$ reduction of the
near-horizon limit $r\rightarrow 0$, the constant 1 in the harmonic
function $H$ can be dropped, and the spacetime then becomes a product
K$_5\times S^5$, where K$_5$ is the generalised Kaigorodov metric in
$D=5$:
\bea
ds_{10}^2&=&Q_1^{-1/2} r^2\, (-K^{-1}\, dt^2 + 
K\, (dx_1 + (K^{-1} -1) dt)^2 + dx_2^2 + dx_3^2)\nn\\
&& + Q_1^{1/2} r^{-2} dr^2 +Q_1^{1/2} d\Omega_4^2\ .
\eea
Compactifying the solution on the $S^5$, with $ds_5^2 = Q_1^{1/2}\,
d\Omega_5^2$, we obtain the five-dimensional Einstein metric
\be
ds_5^2 = Q_1^{1/2}\, \Big( -e^{6\rho}\, dt^2 + e^{-2\rho}\, 
(dx_1+ e^{4\rho}\, dt)^2 + e^{2\rho}\, (dx_2^2 + dx_3^2) +
d\rho^2\Big)\ .\label{d5kaig}
\ee 
This is precisely the generalisation of the Kaigorodov metric to
$D=5$, derived in Appendix A, which is a
solution to $D=5$ gravity with a pure cosmological term $e^{-1}
{\cal L}_5 =R-3\Lambda$ with $\Lambda= -16\, Q_1{-1/2}$.

     We may again consider the limit where the dynamics of the
D3-brane decouples from the bulk.  Note that we have $Q_1=N \ell_p^4$
and $Q_2 = P \ell_p^8$, where $\ell_p=\kappa_{10}^{1/4}$ and $P$ is
the momentum density of the world-volume spatial dimensions.  Note
also that we have $\ell_p^2 = g^{1/2}\, \a'$ where $g$ is the string
coupling constant.  In the limit $\ell_p\rightarrow 0$, with
$U=r/(\sqrt{N}\, \ell_p^2)$ fixed, one has $N\ell_p^4/r^4 >>1$, and
hence the 1 in the harmonic function $H$ can be dropped \cite{mald},
giving rise to the metric
\be
ds_{10}^2 =\ell_p^2\, N^{1/2} \Big(\fft{P}{N^2}\, \fft{dx_1^2}{U^2} +
U^2\, (2dx_1\, dt + dx_2^2 + dx_3^2) +
\fft{dU^2}{U^2} + d\Omega_5^2\Big)\ .
\ee
Thus we see that in this limit, the metric $ds_{10}/\ell_p^2$ is
independent on $\ell_p$, and the system has a fixed wave-momentum
$P$. If $P$ is zero, it reduces to the previously-known AdS$_5\times
S^5$ metric.  When $P$ is instead non-vanishing, we have K$_5\times
S^5$.  Since the generalised Kaigorodov metric K$_5$ is an
infinitely-boosted AdS$_5$, we expect that string theory on this
K$_5\times S^5$ background is dual to $N=4$, $D=4$ Yang-Mills theory
on an infinitely-boosted frame, with constant momentum density
$P$.

        Analogously, we may again consider also the non-extremal
solution with a superimposed gravitational pp-wave,
\bea       
ds_{11}^2 &=& H^{-1/2}(-K^{-1}\, e^{2f}\, dt^2 +
    K(dx_1 + \coth\mu_2\, (K^{-1}-1)dt)^2 + dx_2^2 +
dx_3^2) \nn\\
 && + H^{1/2}(e^{-2f} dr^2 + r^2 d\Omega_5^2)\ ,\nn\\
F_5 &=& \coth\mu_1 (dH^{-1}\wedge d^4x + {*(dH^{-1}\wedge d^4x)}) \ ,
\label{d3nonextm}
\eea
where 
\bea
H=1 + \fft{\kappa_{10}\, k}{r^4}\, \sinh^2\mu_1\ ,\qquad
K=1 + \fft{\kappa_{10}\, k}{r^4}\, \sinh^2\mu_2\ ,\qquad
e^{2f} = 1 -\fft{\kappa_{10}\,k}{r^4}\ .\label{d10hkf}
\eea
The horizon of the non-extremal boosted D3-brane is at
$r_+=\kappa_{10}^{1/4}\, k^{1/4}$.  As for the non-extremal M-branes
discussed previously, the coordinate transformation (\ref{boost}) locally
maps the solution (\ref{d3nonextm}) to the unboosted one, where $K=1$.
 
      First, let us consider the double-dimensional reduction on the
world-volume coordinates $(x_1, x_2, x_3)$.  This gives rise to
a 2-charge non-extremal isotropic black hole in $D=7$.  The relevant
Lagrangian that describes this solution is \label{lpsol,cjlp1}
\be
e^{-1}\,\kappa_7^2\, {\cal L}_7 =
R-\ft12(\del \vec \phi)^2 -\ft1{44} e^{\vec a_{34}\cdot \vec \phi}\, 
(F_{\2 34})^2 -\ft14 e^{\vec a_{12} \cdot \vec \phi}\, (F_{\2 12})^2\ ,
\ee
where $\vec\phi=(\phi_1, \ldots, \phi_4)$ and
\bea 
\vec a_{12}&=&(1, 3/\sqrt7, -1/\sqrt{21}, -1/\sqrt{15})\ ,\nn\\
\vec a_{34}&=&(-\ft12, -3/(2\sqrt7), 4/\sqrt{21}, 4/\sqrt{15})\ .
\eea
The seven-dimensional gravitational constant is given by $\kappa_7^2 =
\kappa_{10}^2/(L_1L_2 L_3)$, where $L_i$ is the period of the
coordinate $x_i$.

       The seven-dimensional non-extremal 2-charge black hole is given by
\bea
&&ds_{7}^2=-(H\,K)^{4/5}\, e^{2f}\, dt^2 +
(H\,K)^{1/5}(e^{-2f}\, dr^2 + r^2 d\Omega_5^2)\ ,\nn\\
&& \vec \phi =\ft12 \vec a_{34}\, \log H + 
\ft12 \vec a_{12}\, \log K\ ,\nn\\
&& A_{\1 34} = \coth\mu_1\, H^{-1}\, dt\ , \qquad 
A_{\1 12} = \coth\mu_2\, K^{-1}\, dt\ .\label{d7bh}
\eea
It is straightforward to see that the entropy of the black hole is
\bea
S&=& \fft{\hbox{Area}}{4\kappa_7^2}\ ,\nn\\
&=& \fft{L_1L_2 L_3}{\kappa_{10}^{3/4}}\, \Omega_5\, k^{5/4}\,
\cosh\mu_1 \, \cosh\mu_2\ .\label{d7bhent}
\eea

       Let us now consider instead the $S^5$ reduction of the boosted
D3-brane.  In the near-extremal limit, the constant 1 in $H$ can be
dropped near the horizon, and hence the spacetime becomes a product
$M_5\times S^5$.  The metric of the internal 5-sphere is $ds_5^2 =
\kappa_{10}^{1/2}\, (k\sinh^2\mu_1)^{1/2}\, d\Omega_5^2$, 
and so its radius is $R_5=
\kappa_{10}^{1/4}\, (k\sinh^2\mu_1)^{1/4}$.  It follows that the
five-dimensional gravitational constant is given by
\be
\kappa_5^2 = \fft{\kappa_{10}^2}{V_{S^5}} = 
\fft{\kappa_{11}^{3/4}}{(k\sinh^2\mu_1)^{5/4}\, \Omega_5}\ .
\ee

     Implementing the $S^5$ reduction, we obtain the five-dimensional
generalisation of the Carter-Novotn\'y-Horsk\'y metric, derived in
Appendix C:
\bea
ds_5^2 &=& (\kappa_{10}\, k\, \sinh^2\mu_1)^{-1/2}\,  r^2\, 
\Big(-K^{-1}\, e^{2f}\, dt^2 + K(dx_1 +  
\coth\mu_2\, (K^{-1} -1)\, dt)^2\nn\\
&& + dx_2^2 +dx_3^2\Big)  + 
(\kappa_{10}\, k\, \sinh^2\mu_1)^{2/3}\, e^{-2f}\, 
\fft{dr^2}{r^2} \ .\label{d5nonext}
\eea
Again this solution is still Einstein, but it is no longer homogeneous.
In the asymptotic region $r\rightarrow \infty$, the metric approaches the
generalised homogeneous Kaigorodov metric.  The entropy of the
metric (\ref{d5nonext}) is given by
\bea
S &=& \fft{\hbox{Area}}{4\kappa_5^2}\nn\\
  &=& \fft{L_1L_2 L_3}{\kappa_{10}^{3/4}}\, \Omega_4\, k^{5/4}\,
\sinh\mu_1 \, \cosh\mu_2\ ,\label{d5dent}
\eea
which agrees with (\ref{d7bhent}) in the near-extremal limit $\mu_1
>>1$.  In the case where the wave is absent, the entropy and
temperature satisfy the ideal-gas relation $S\sim T^3$; however, the
presence of the wave alters the relation, and it becomes $S\sim
T^{1/3}$.  As in the previous cases, the $\mu_2$ dependence of the
entropy density is a natural consequence of the Lorentz contraction
along the direction of the boost on the world-volume, implying that
the entropy density is enlarged by the factor $\cosh\mu_2$, which is
the $\gamma$-factor of the Lorentz boost.  Thus in the near extremal
regime, the system can be modelled by an ideal gas in a boosted frame
in four dimensional spacetime.  When the system is highly boosted, but
with the momentum density $k\, e^{2\mu_2}$ held fixed, then it gives
rise to the entropy of the near-extremal 2-charge black hole in $D=7$.
This is consequence of the correspondence that type IIB supergravity
on K$_5\times S^5$ is dual to the $D=4$, $N=4$ Yang-Mills theory in an
infinitely-boosted frame, with constant momentum density, where K$_5$
is the five-dimensional generalisation of the Kaigorodov metric.

\section{Dyonic string with pp-wave}

     Lower dimensional examples such as AdS$_3$ and AdS$_2$ also arise
as the near-horizon limits of supergravity $p$-branes.  AdS$_3\times
S^3$ is the near horizon of the dyonic string in $D=6$.  When a wave
is propagating on the worldsheet of the string with momentum density
$P$, it gives rise to a 3-charge non-dilatonic black hole in $D=5$.
The metric of the spacetime AdS$_3\times S^3$ plus a wave is given by
\be
ds_6 =\ell_p^2\,\sqrt{N_1N_2}\Big(
\fft{P}{N_1N_2}\, dx^2 + \fft{2r^2}{N_1 N_2\, \ell_p^4}\, 
dx\, dt + \fft{dr^2}{r^2} + d\Omega_3^2 \Big)\ ,
\ee
where $\ell_p=\kappa_6^{1/2}$, and $N_1$, $N_2$ are the electric and
magnetic charges of the dyonic string, and $P$ is the momentum density
of the wave. Note that if $P=0$, the three-dimensional metric obtained
by dimensional reduction on $S^3$ is AdS$_3$, in horospherical
coordinates.  When $P$ is non-vanishing, owing to the degeneracy of
three-dimensional gravity,\footnote{In three dimensions the Riemann
tensor is characterised completely by the Ricci tensor, and
consequently any three-dimensional Einstein metric with negative
cosmological constant is locally equivalent to AdS$_3$.} the metric is
still locally AdS$_3$, but the global structure is different; this is
the $D=3$ case of the generalised Kaigorodov metrics obtained in
Appendix A.  This metric is equivalent to the BTZ black-hole metric
\cite{btz}, in the extremal limit where $J=M\, \ell$ (here
$-2\ell^{-2}$ is the cosmological constant).  The boundaries of the
AdS$_3$ and three-dimensional Kaigorodov (or extremal BTZ) metrics are
different: In horospherical coordinates, the boundary of AdS$_3$ is a
two-dimensional Minkowski spacetime, whilst in the above metric, the
boundary is two-dimensional a spacetime in the infinite-momentum
frame.  The gravitational decoupling limit is $\ell_p\rightarrow 0$,
while keeping $U=r/\sqrt{N_1\,N_2\,\ell_p^4}$ fixed.

       The different global structures of the horospherical AdS$_3$
and the spacetime arising in the case where there is a pp-wave can
also be seen from the entropy/temperature relation in the
near-extremal regime.  When there is a wave propagating on AdS$_3$, we
have $S\sim T^0$.  On the other hand when the wave is absent, we have
$S\sim T$, which is the ideal-gas relation in one dimension.  When the
extra Kaluza-Klein charge is included, the new parameter $\mu_2$ in
(\ref{d5dent}) is again the natural consequence of the associated
Lorentz contraction of the volume, implying the dilation of the
entropy density by a factor of $\gamma=\cosh\mu_2$.  Thus the
microscopic entropy of the classical solution with the presence of the
wave can be modelled by a dilute massless gas in a two-dimensional
spacetime in a boosted frame.  In particular the three-charge BPS black
hole in $D=5$ has non-vanishing entropy, which corresponds
microscopically to a dilute gas in an infinitely-boosted frame, but
with the momentum density held fixed.

      AdS$_2$ spacetime arises in supergravity as the near-horizon
geometry of the extremal Reissner-Nordstr{\oo}m-type black hole in $D=5$
and $D=4$.  Its boundary is one dimensional, and hence there can be no
propagating wave.

\section{Conclusions and Discussion}

     In this paper, we have studied the three cases of single-charge
non-dilatonic $p$-branes, namely the M2-brane and M5-brane of
M-theory, and the D3-brane of the type IIB string, in the presence of
a gravitational pp-wave propagating in the world-volume.  When
dimensionally reduced on all the spatial world-volume coordinates,
these configurations give rise to 2-charge black holes in $D=9$, 6 and
7.  One of the charges comes from the original 4-form or 5-form
antissymmetric tensor charge in $D=11$ or $D=10$, while the other is
carried by a Kaluza-Klein vector.  If the configuration is
non-extremal, the effect of the inclusion of the pp-wave is locally
equivalent to a Lorentz boost on the world-volume of the $p$-brane,
but for extremal configurations the corresponding boost would be
singular, with a boost velocity equal to the speed of light.

   The near-horizon structure of the M2-brane, M5-brane or D3-brane
with a pp-wave is of a product form, $M_4\times S^7$, $M_7\times S^4$
or $M_5\times S^5$, where in the extremal case $M_n$ is the
$n$-dimensional generalisation K$_n$ of the four-dimensional
Kaigorodov metric.  In the non-extremal case $M_n$ is the
$n$-dimensional generalisation C$_n$ of the four-dimensional
Carter-Novotn\'y-Horsk\'y metric.  The metrics K$_n$, which we
construct in Appendix A, are homogeneous Einstein metrics.  The
metrics C$_n$, which we construct in Appendix C, are inhomogeneous
Einstein metrics.  Since the local structure of the non-extremal
$p$-branes is the same whether or not there is a pp-wave present,
there are only global differences between the structures of the
generalised Carter-Novotn\'y-Horsk\'y metrics that correspond to the
$p$-branes with and without the pp-wave.  On the other hand in the
extremal case there is no non-singular boost that can relate the
solution with the pp-wave to the one without, and for this reason the
generalised Kaigorodov metric K$_n$ is not even locally the same as
the AdS$_n$ metric which would arise in the $M_n\times$Sphere product
in the near-horizon limit of the extremal $p$-brane with no pp-wave.
In Appendix B we construct the Killing vectors and Killing spinors on
the generalised Kaigorodov metrics. In particular, we find that K$_n$
admits just $1/4$ of the maximal number of Killing spinors that occur
on AdS$_n$.

    We have argued that by considering the extremal M2-brane, M5-brane
or D3-brane in the presence of a pp-wave, the conjectured relations
between supergravity in AdS$_n$ and conformal field theories on its
boundary can be generalised to relations between supergravity on the
Kaigorodov-type metric K$_n$ and a CFT on its boundary.  Specifically,
this boundary is related to the usual AdS$_n$ boundary by an infinite
Lorentz boost, and so the expected conformal field theories will now
be related to those of the AdS$_n$ backgrounds by a singular passage
to the infinite-momentum frame.  The decoupling limit, where the
gravitational constant is sent to zero, requires holding the momentum
density of the wave fixed.  This correspondence is consistent with the
supersymmetries of the two theories.  In the supergravity picture, the
Kaigorodov metric preserves just $1/4$ of the supersymmetry.  On the
rest-frame CFT side, the superconformal invariance enhances
supersymmetry by doubling the number of conserved supercharges.
In the infinitely-boosted frame, the non-vanishing momentum implies
that half of the original supersymmetry, as well as the superconformal
symmetry is also broken. Thus it follows that the theory has 
just $1/4$ of the conserved supercharges.

    We also considered the macroscopic and microscopic entropies of
2-charge black holes in $D=9$, 7 and 6 in their near-extremal
regimes.  We showed that these entropies are related to those of the
corresponding single-charge black holes by factors that can be
accounted for as Lorentz contractions of the world-volume along the
direction of the boost that relates the solutions with and without the
pp-wave.  Consequently, the microscopic entropy of such a
near-extremal black hole can be described in terms of a boosted dilute
gas of massless particles on the world-volume of the original
$p$-brane. In other words, we have $S=\cosh\mu_2\, S_{\rm dilute\,\,
gas}$, for any boost parameter $\gamma=\cosh\mu_2$.  (We also showed
that the macroscopic entropy of the 2-charge black holes in their
near-extremal regimes can also be calculated in the associated
generalised Carter-Novotn\'y-Horsk\'y metrics, obtained by
dimensionally-reducing the original $p$-brane plus pp-wave solutions
on the foliating spheres of the transverse space.)  If we consider
a dilute gas on the world-volume of the $p$-brane in a highly-boosted frame,
but with the momentum density fixed, then the entropy and temperature
satisfy a relation $S\sim T^{1/(\td d -2)}$, where $\td d$ is the
dimension of the foliating sphere of the space transverse to the
$p$-brane.  This observation
suggests that the co-dimension $(\td d +1)$ of the $p$-brane seems to
be encoded in the CFT theory on an infinitely-boosted frame with
constant momentum density.

        It is worth remarking that the Kaigorodov metric in $D=4$
arises as the near-horizon geometry of the intersection of an M2-brane and
a pp-wave, which is the oxidisation of a ten-dimensional 
D0-brane.  This may lead to a 
connection between the CFT and the M(atrix) model on an AdS background.

       We should like to conclude with proposals for future study,
beyond the scope of this paper. In the CFT in the infinitely-boosted
frame, the only surviving states are those with purely transverse
polarisations, and their correlation functions should reflect this
fact. Note also that in the rest-frame CFT the correlation functions
are usually calculated in a Euclideanised spacetime, 
while in the current context the Minkowskian nature
of the spacetime becomes crucial.  On the dual side, this information
about the field excitations is encoded in the perturbations of the
background of the Kaigorodov-type metrics (the near-horizon region of
BPS $p$-branes with pp-waves). It is thus of interest to address 
these issues 
both in the CFT and on the gravity side, in order to shed further
light on the nature of the correspondence in the infinitely-boosted
frame.

\bigskip\bigskip
\noindent{\Large{\bf Acknowledgments}}
\bigskip

    We are very grateful to Gary Gibbons and Steven Siklos for
discussions about the Kaigorodov metric, to Igor Klebanov and Juan
Maldacena for discussions on CFT in boosted frames, and decoupling
limits, and Glen Agnolet, Mike Duff, Zachary Guralnik, Randy Kamien,
Tom Lubensky and Akardy Tseytlin for discussions.



\section*{Appendices}

\appendix

\section{$D$-dimensional generalisation of the Kaigorodov metric}

   Let us consider the following family of metrics in $D=n+3$
dimensions:
\be
ds^2 = -e^{2a\rho} \, dt^2 + e^{2b\rho}\, (dx+ e^{(a-b)\rho}\, dt)^2 
  + e^{2c\rho}\, dy^i\, dy^i + d\rho^2\ ,\label{kaig}
\ee
where $a$, $b$ and $c$ are arbitrary constants.  It is easily seen
that these encompass the metrics that we obtained in
this paper by the spherical dimensional reduction of the extremal
M2-brane, M-5-brane and D3-brane with pp-waves.  Choosing the natural
orthonormal basis
\be
e^0= e^{a\rho}\, dt\ ,\qquad e^1= e^{b\rho}\, (dx + e^{(a-b)\rho}\,
dt)\ ,\qquad e^2 = d\rho\ ,\qquad e^i = e^{c\rho}\, dy^i\ ,
\ee
where $3\le i\le n+2$, we find that the torsion-free spin connection,
defined by $de^a = -\omega^a{}_b\wedge e^b$,
$\omega_{ab}=-\omega_{ba}$, is given by
\bea
&&\omega_{01} = \ft12(a-b)\, e^2\ ,\qquad \omega_{0i} = 0\ ,\qquad
\omega_{02} = -a\, e^0 + \ft12(a-b)\, e^1\ ,\nn\\
&&\omega_{12} = b\, e^1 + \ft12(a-b)\, e^0\ ,
\qquad \omega_{1i} = 0\ ,\qquad \omega_{2i} = -c\, e^i\ ,\qquad
\omega_{ij} = 0\ .\label{spincon}
\eea
It is immediately evident from this that the metrics are homogeneous,
since all orthonormal components of the spin connection, and hence of
the curvature, are constants.  Consequently, all curvature invariants
are constants.  We find that the curvature 2-forms, defined by
$\Theta_{ab} = d\omega_{ab} + \omega_a{}^c\wedge \omega_{cb}$, are
given by
\bea
&&\Theta_{01} = \ft14(a+b)^2\, e^0\wedge e^1\ ,\qquad
\Theta_{02} = \ft14(a^2 + 6 a b -3 b^2)\, e^0\wedge e^2 - b(a-b)\, 
e^1\wedge e^2\ ,\nn\\
&&
\Theta_{0i} = a c\, e^0\wedge e^i - \ft12 c(a-b)\, e^1\wedge e^i\ ,
\qquad \Theta_{ij} = -c^2 \, e^i\wedge e^j\ ,
\nn\\
&&
\Theta_{12} = -\ft14(a^2 -2ab + 5 b^2)\, e^1\wedge e^2 - b(a-b)\,
e^0\wedge e^2\ ,\nn\\
&&
\Theta_{1i} = -b c\, e^1\wedge e^i - \ft12 c(a-b)\, e^0\wedge e^i\ ,
\qquad
\Theta_{2i} = -c^2\, e^2\wedge e^i\ .\label{curv2forms}
\eea
From this, we find that the Ricci tensor has the vielbein components
\bea
&&
R_{00} = \ft12 a^2 + 2 a b -\ft12 b^2 + n\, a c\ ,\qquad 
R_{11} = -\ft12 a^2 -\ft32 b^2 -n\, b c\ ,\nn\\
&&
R_{22} = -\ft12(a+b)^2 - n\, c^2 \ ,\qquad
R_{ij} = -(a + b + n\, c)\, c\, \delta_{ij}\ ,\nn\\
&&
R_{01} = -\ft12 (a-b)\, (2\, b + n\, c) \ .\label{ricci}
\eea

    Requiring that the metrics be Einstein, namely that the vielbein
components of the Ricci tensor obey $R_{ab}= \Lambda\, \eta_{ab}$, 
we find that there are exactly two inequivalent solutions, {\it viz}.
\bea
{\rm AdS}_{n+3}: && a=b=c= 2L\ ,\label{adsconsts}\\
{\rm K}_{n+3}: && a=(n+4)\, L\ ,\qquad b=-n\, L\ ,\qquad c = 2\, L\ ,
\label{kaigconsts}
\eea
where $L=\ft12\sqrt{-\Lambda/(n+2)}$ and the cosmological constant
$\Lambda$ is negative.  The first family of Einstein metrics
corresponds to anti-de Sitter spacetime in $D=n+3$, while the second
family corresponds to $D=n+3$ homogeneous Einstein metrics that
generalise the Kaigorodov metric of four dimensions.  Note that this
second family, of generalised Kaigorodov metrics, can be written in
the form
\be
ds^2 = e^{-2nL\rho}\, dx^2 +  e^{4L\rho}\, (2 dx\, dt + dy^i\, dy^i)
+ d\rho^2\ .\label{kmet}
\ee
Substituting the constants $a$, $b$ and $c$ given by
(\ref{kaigconsts}) into (\ref{curv2forms}), we find that the curvature
2-forms $\Theta_{ab}$ for the generalised Kaigorodov metrics can be
written in terms of the Weyl 2-forms $C_{ab}=\ft12C_{abcd}\, e^c\wedge
e^d$, where $C_{abcd}$ is the Weyl tensor, as follows
\be
\Theta_{ab} = -4L^2 \, \eta_{ac}\, \eta_{bd}\, e^c\wedge e^d + C_{ab}
\ .\label{thetc}
\ee
Here, the Weyl 2-forms are given by
\bea
&&C_{12} = -C_{02} = n\, \mu\, (e^0-e^1)\wedge e^2\ ,\nn\\
&&C_{0i} = -C_{1i} = \mu\, (e^0 -e^1)\wedge e^i\ ,
\label{weyltensor}\\
&&C_{01}= C_{ij}= C_{2i} = 0\ ,\nn
\eea
where $\mu = 2(n+2)\, L^2$.  Thus we see that the Weyl tensor is
non-zero in the generalised Kaigorodov metrics, although it has a a
rather simple structure.  Note that the vielbein combination $e^0-e^1$
that appears in all the non-vanishing Weyl 2-form components is simply
given by $e^0-e^1 = -e^{b\rho}\, dx$.  The vector dual to the 1-form
$-(e^0-e^1)$ is simply $K_{(0)}={\del}/{\del t}$.  This is a null
Killing vector, and a zero eigenvector of the Weyl tensor, satisfying
$C_{abcd}\, K^d_{(0)}=0$.  The generalisations of the Kaigorodov
metric that we have obtained here can be interpreted as describing
gravitational waves propagating in an anti-de Sitter spacetime
background.  (This is discussed for four-dimensional Kaigorodov metric
itself in \cite{pod}.)

     In the case of four dimensions, the Kaigorodov metric is of type
N in the Petrov classification (see, for example, \cite{grbook}).  If
we define the dual of the Weyl tensor by $\wtd C_{abcd} = \ft12
\epsilon_{abef}\, C^{ef}{}_{cd}$, and thence the complex Weyl tensor
$W_{abcd} \equiv C_{abcd} + \im\, \wtd C_{abcd}$, then it is easily
seen that we can write $W_{abcd}$ in the null form
$W_{abcd}=-4V_{ab}\, V_{cd}$, where the 2-form $V=\ft12 V_{ab}\,
e^a\wedge e^b$ is given by
\be
V= \sqrt{\fft{-\Lambda}{8}}\, (e^0-e^1)\wedge (e^2-\im\, e^3)\ .
\ee
The null Killing vector $K_{(0)}={\del}/{\del t}$ is the quadruple
Debever-Penrose null vector of the type-N Weyl tensor \cite{pod}.

     In three dimensions, the Kaigorodov metric becomes simply $ds^2 =
dx^2 + 2e^{4L\rho}\,dx\, dt + d\rho^2$.  This can be seen to be
equivalent to the extremal limit of the BTZ black-hole metric
described in \cite{btz}, where the angular momentum $J$ and mass $M$
are related by $J=M\, \ell$, and $-2\ell^{-2}$ is the cosmological
constant.

   The family of Einstein metrics in (\ref{adsconsts}), by contrast,
corresponds to the AdS metrics, with
\bea
ds^2 &=& e^{4L\rho}\, (-dt^2 + (dx+dt)^2 + dy^i\, dy^i) + d\rho^2\ , \nn\\
&=& e^{4L\rho}\, (-dt^2 + {dx'}^2 + dy^i\, dy^i) + d\rho^2\ ,
\label{adsmet}
\eea
where in the second line we have made the coordinate redefinition
$x'=x+t$ to give the metric its standard horospherical form.  
The Weyl tensor of course vanishes
for this solution, and so then the curvature 2-forms are simply given by
(\ref{thetc}) with $C_{ab}=0$.

    Note that the AdS metrics can be obtained from the generalised
Kaigorodov metrics by taking an appropriate singular limit.  If we
make the redefinitions
\be
x\longrightarrow \ft{\lambda}{\sqrt2}\, (x+t)\ ,\qquad
t\longrightarrow \ft{1}{\lambda\sqrt2}\, (x-t)\ ,
\ee
to the coordinates $x$ and $t$ appearing in the generalised Kaigorodov
metrics (\ref{kmet}), and then send the constant $\lambda$ to zero, we
find that (\ref{kmet}) limits to the AdS metric given in the second
line of (\ref{adsmet}).  Of course the fact that a singular limit is
involved in this procedure means that the AdS and Kaigorodov metrics
are inequivalent, as evidenced, for example, by the fact that the AdS
metrics have vanishing Weyl tensor while the generalised Kaigorodov
metrics do not.

\section{Killing vectors and spinors in the Kaigorodov metrics}

     It is easily seen by inspection that the following are $(\ft12
n^2 + \ft32 n + 3)$ Killing vectors of the generalised Kaigorodov metrics:
\bea
&&K_{(0)} = \fft{\del}{\del t}\ ,\qquad
K_{(x)} = \fft{\del}{\del x}\ ,\qquad
K_{(i)} = \fft{\del}{\del y^i}\ ,\nn\\
&&L_{(i)} = x\, \fft{\del}{\del y^i} - y^i\, \fft{\del}{\del t}\ ,
\qquad L_{(ij)} = y^i\, \fft{\del}{\del y^j} - y^j\, \fft{\del}{\del
y^i}\ ,\label{kvs}\\
&&J = \fft{\del}{\del \rho} - a\, t\, \fft{\del}{\del t} - b\, x\, 
\fft{\del}{\del x} - c\, y^i\, \fft{\del}{\del y^i}\ ,
\eea
where $a$, $b$ and $c$ are given by (\ref{kaigconsts}).
The $K_{(0)}$, $K_{(x)}$ and $K_{(i)}$ Killing vectors mutually
commute, and the rest of the algebra of the Killing vectors is
\bea
&& {[} J, K_{(0)} {]} = a\, K_{(0)}\ ,\qquad
{[} J, K_{(x)} {]} = b\, K_{(x)}\ ,\qquad
{[} J, K_{(i)} {]} = c\, K_{(i)}\ ,\nn\\
&& 
{[} J, L_{(i)} {]} = (a-c)\, L_{(i)}\ ,\qquad
{[} J, L_{(ij)} {]} = 0\ ,\nn\\
&& {[} K_{(x)}, L_{(i)} {]} = K_{(i)} \ ,\qquad
{[} L_{(i)}, K_{(j)} {]} = \delta_{ij}\, K_{(0)}\ ,\qquad
{[} L_{(ij)}, K_{(k)} {]} = -\delta_{ik}\, K_{(j)} + \delta_{jk}\,
K_{(i)}\ ,\nn\\
&& {[} L_{(i)}, L_{(j)} {]} =0\ ,\qquad
{[} L_{(ij)}, L_{(k)} {]} = -\delta_{ik}\,L_{(j)} + \delta_{jk}\,
L_{(i)}\ ,\nn\\
&& {[} L_{(ij)}, L_{(k\ell)} {]} = 
-\delta_{ik}\, L_{(j\ell)} + \delta_{jk}\, L_{(i\ell)}
-\delta_{j\ell}\, L_{(ik)} + \delta_{i\ell}\, L_{(jk)}\ .
\label{comrel}
\eea
Since the metrics are homogeneous, these symmetries act transitively
on the spacetimes.   In the four-dimensional case, we have the
previously-known five-dimensional group of symmetries on the Kaigorodov
spacetime \cite{kaig}.

     The Killing spinor equation in a $D$-dimensional spacetime is
\be
D_\mu\, \epsilon^\pm = \pm \sqrt{-\fft{\Lambda}{D-1}}\, 
\Gamma_\mu\, \epsilon^\pm\ ,
\label{kseqn}
\ee
where $D_\mu = \del_\mu + \ft14 \omega_\mu^{ab}\, \Gamma_{ab}$.
Thus for the generalised Kaigorodov metrics described above, the
equation becomes $D_\mu\, \epsilon^\pm = \pm \ft12 c \, 
\Gamma_\mu\, \epsilon^\pm$.

     It is instructive first to consider the integrability conditions
for the existence of Killing spinors.  The simplest of these is the
2'nd-order condition that results from taking a commutator of the
derivatives $\bar D_\m^\pm \equiv D_\mu \mp \ft12 c\, \Gamma_\mu$
arising in the Killing spinor equation $\bar D_\mu^\pm \, \ep^\pm =
0$.  Thus we obtain the condition
\be
H_{\mu\nu}\, \ep^\pm \equiv 
4 {[} \bar D_\mu^\pm, \bar D_\nu^\pm {]} \,\ep^\pm = R_{\mu\nu\rho\sigma}\, 
\Gamma^{\rho\sigma}\, \ep^\pm +  2 c^2\, \Gamma_{\mu\nu}\, \ep^\pm =0\ . 
\label{2ndorder} 
\ee
Note that the quantity $H_{\mu\nu}$ can be written simply as
$H_{\mu\nu} = C_{\mu\nu\rho\sigma}\, \Gamma^{\rho\sigma}$, where
$C_{\mu\nu\rho\sigma}$ is the Weyl tensor, discussed in the previous
section.  Upon substitution of the Riemann tensor, (\ref{2ndorder})
gives algebraic conditions on the Killing spinors $\ep^\pm$, in the
form of projection operators formed from the $\Gamma$ matrices.  These
are necessary conditions for the existence of Killing spinors, and in
many cases they are also sufficient.  (See \cite{wvn} for a discussion
of higher-order integrability conditions for Killing spinors.)
However, as we shall see, for the generalised Kaigorodov metrics these
2'nd-order integrability conditions are not in fact sufficient.
Before proceeding to study (\ref{2ndorder}), therefore, let us present
also the 3'rd-order integrability condition that follows by taking a
further derivative of (\ref{2ndorder}), and using the original Killing
spinor equation again.  Thus we obtain
\be
H_{\lambda\mu\nu}^\pm\, \ep^\pm \equiv 
(\nabla_\lambda\, R_{\mu\nu\rho\lambda}) \Gamma^{\rho\sigma}\, \ep^\pm
\mp 2 c\, R_{\mu\nu\lambda\rho}\, \Gamma^\rho\, \ep^\pm \pm 2c^3\,
(g_{\nu\lambda}\, \Gamma_\mu -g_{\mu\lambda}\, \Gamma_\nu)\, \ep^\pm =0
\ .
\label{3rdorder}
\ee

    From (\ref{curv2forms}), and substituting the solution for $a$,
$b$ and $c$ in (\ref{kaigconsts}), we find that the
quantities $H_{\mu\nu}$ in the 2'nd-order integrability condition 
(\ref{2ndorder}) are given by
\bea
&&H_{02} = 2n(n+2)\, L^2 \, (\Gamma_{02} + \Gamma_{12})
\ ,\qquad H_{0i} = -2(n+2)\, L^2\, (\Gamma_{0i} + \Gamma_{1i})\ ,\nn\\
&&
H_{12} =-2n(n+2)\,L^2\,  (\Gamma_{02} + \Gamma_{12})\ ,
\qquad H_{i1} = 2(n+2)\,L^2\,  
(\Gamma_{0i} + \Gamma_{1i}) \ ,\label{2ndorderh}\\
&& H_{01} = 0\ ,\qquad H_{ij} = 0\ ,\qquad H_{2i} = 0\ .
\eea
Thus the integrability conditions $H_{\mu\nu}\, \ep^\pm = 0$ imply that
$\epsilon^\pm$ must satisfy
\be
\Gamma_{01}\, \ep^\pm = \ep^\pm\ .\label{2con}
\ee
One might be tempted to think that this were the only condition, in
which case the Killing spinors would preserve half of the maximal
supersymmetry.  However, as foreshadowed above, higher-order
integrability can place further constraints in certain cases, and in
fact the present example is one such.  It suffices to consider just
one special case among the 3'rd order conditions implied by
(\ref{3rdorder}).  Consider, for example, the components $H_{10i}^\pm$
of $H_{\lambda\mu\nu}^\pm$.  The covariant derivative of the Riemann
tensor can be evaluated using the general expression $\nabla_\mu\,
V_\nu = \del_\mu\, V_\nu + (\omega_\nu{}^\rho)_\mu\, V_\rho$, and
after some algebra we find that $\nabla_1\, R_{0i\rho\sigma}\,
\Gamma^{\rho\sigma}= 8(n+2)\, L^2\, \Gamma_{2i}$.  Substituting into
(\ref{3rdorder}), we therefore find that $H_{10i}^\pm = 8(n+2)\, L^2\, 
(\Gamma_{2i} \pm \Gamma_i)$, implying that $\ep^\pm$ must also satisfy
the condition
\be
\Gamma_2\, \ep^\pm = \pm\ep^\pm \ ,\label{3con}
\ee
in addition to (\ref{2con}). In principle we should examine all the
components of $H_{\lambda\mu\nu}^\pm$, but the upshot is that no
further conditions result.  The simplest way to prove this is by
moving now to an explicit construction of Killing spinors that satisfy
the two conditions (\ref{2con}) and (\ref{3con}).

Substituting the spin connection (\ref{spincon}) into this, we find
that the Killing spinors must satisfy the following system of
equations:
\bea
\fft{\del\epsilon^\pm}{\del t} + \ft14 (a+b)\, e^{a\, \rho}\,
(\Gamma_{02} + \Gamma_{12}) \, \epsilon^\pm &=&\pm  \ft12
 c\, e^{a\, \rho}\, (\Gamma_0 + \Gamma_1)\, \epsilon^\pm\ ,\nn\\
\fft{\del\epsilon^\pm}{\del x}+ e^{b\, \rho}\, \Big(\ft12b\, 
\Gamma_{12} -\ft14
(a-b)\, \Gamma_{02}\Big)\, \epsilon^\pm &=& \pm\ft12 c\, e^{b\, \rho}\,
\Gamma_1\, \epsilon^\pm\ ,\nn\\
\fft{\del\epsilon^\pm}{\del y^i} -\ft12 c\, e^{c\, \rho}\,
\Gamma_{2i}\, \epsilon^\pm &=& \pm\ft12 c\, e^{c\, \rho}\, \Gamma_i\,
\epsilon^\pm\ ,\nn\\
\fft{\del\epsilon^\pm}{\del \rho} -\ft14(a-b)\, \Gamma_{01}\,
\epsilon^\pm &=& \pm\ft12 c\, \Gamma_2\, \epsilon^\pm\ .
\eea
It is easily seen from these equations and from (\ref{kaigconsts})
that the Killing spinors are given by $\epsilon^\pm =
e^{\fft12(n+4)L\, \rho}\, \epsilon_0^\pm$, where $\epsilon_0^\pm$ is
any constant spinor that satisfies the conditions
\be
\Gamma_2\, \epsilon_0^\pm = \pm \epsilon_0^\pm\ ,\qquad
\Gamma_{01}\, \epsilon_0^\pm = \epsilon_0^\pm\ ,
\ee
implying that $\ep^\pm$ satisfies the conditions (\ref{2con}) and
(\ref{3con}) that followed from integrability.  Thus by combining the
necessary conditions coming from integrability with the sufficient
conditions coming from the explicit solutions, we conclude that we
have found the general solutions for the Killing spinors in the
generalisation of the Kaigorodov spacetime, and that they preserve
$1/4$ of the supersymmetry.

\section{Generalisations of Carter-Novotn\'y-Horsk\'y metrics}

   The form of the metrics that arise in the spherical reduction of the 
non-extremal $p$-brane plus wave solutions is
\be
ds^2 = c_1\, e^{2{\td d}\rho/d}\, \Big(- K^{-1}\, e^{2f}\, dt^2 + K\, (dx_1 +
\coth\mu_2\, (K^{-1}- 1) dt)^2 + dy^i\, dy^i \Big) + c_2\, e^{-2f}\,
d\rho^2\ ,\label{nonextmet}
\ee
(see, for example, (\ref{d4nonext}) or (\ref{d7nonext})), where
$e^{2f} = 1-k\, e^{-\td d \rho}$ and $K = 1 + k\, \sinh^2\mu_2\,
e^{-\td d\rho}$.  (We are setting the gravitational constants $\kappa$
to unity here, for convenience.) Let us consider two cases.  First,
when the charge for the harmonic function $K$ associated with the wave
is zero, so that $K=1$, we have $\mu_2=0$ and hence the metric
(\ref{nonextmet}) becomes
\be
ds^2 = c_1\,  e^{2{\td d}\rho/d}\, \Big(- e^{2f}\, dt^2 + dx_1^2 +
dy^i\, dy^i \Big) + c_2\, e^{-2f}\, d\rho^2\ .\label{nonextnowavemet}
\ee

    Now, consider instead the case where the charge associated with the
wave is non-zero.  Let us make the following Lorentz boost on the
coordinates $(t,x_1)$:
\bea
     x_1 &=& x_1' \, \cosh\mu_2 + t'\, \sinh\mu_2\ ,\nn\\
      t' &=& x_1'\, \sinh\mu_2 + t'\, \cosh\mu_2\ .\label{boost}
\eea
Note that in terms of the velocity $v$ for the boost along $x_1$, we
have simply
\be
x_1 = \gamma\, (x_1' - v\, t')\ ,\qquad t = \gamma\, (t' - v\, x_1')
\ ,\label{lorboost}
\ee
where 
\be
v=\tanh\mu_2\ ,\qquad \gamma=(1-v^2)^{-1/2}= \cosh\mu_2\ .\label{gamma}
\ee
After simple algebra, we find that the metric (\ref{nonextmet})
becomes
\be
ds^2 = c_1\,  e^{2{\td d}\rho/d}\, \Big(- e^{2f}\, d{t'}^2 + d{x_1'}^2 +
dy^i\, dy^i \Big) + c_2\, e^{-2f}\, d\rho^2\ ,\label{nonextwavemet}
\ee
which is identical in form to the previously-obtained metric
(\ref{nonextnowavemet}).\footnote{Of course the Lorentz boost
(\ref{boost}) can equally well be applied not only to the
spherically-reduced metrics considered in this appendix, but also to
the original non-extremal $p$-branes with superimposed pp-waves.  In
fact {\it any} non-extremal solution with a superimposed pp-wave can
be mapped by the transformation (\ref{boost}) into a solution where
the wave momentum vanishes and hence the associated harmonic function
$K$ becomes just the identity.  A particularly striking example is
when a non-extremal black hole supported only by a charge for a
Kaluza-Klein vector is oxidised back to the higher dimension.  In this
case, the coordinate transformation (\ref{boost}) maps the
higher-dimensional wave metric into a purely Minkowski metric. An
example of this is the non-extremal D0-brane in $D=10$, which, after
oxidation to a wave in $D=11$, can then be mapped into Minkowski
spacetime.  Note that this cannot be done in the extremal limit,
since the boost transformation (\ref{boost}) then becomes singular.}
Note that the Lorentz boost (\ref{boost}) is a valid coordinate
transformation only if the coordinate $x_1$ is not periodic, but
instead ranges over the entire real line.  Thus it is only when $x_1$
is non-periodic that the effect of the Kaluza-Klein charge can be
``undone'' by the Lorentz boost (\ref{boost}).

   If $x_1$ is non-periodic, the spherical reductions of the
non-extremal pure M2-brane, M5-brane and D3-brane are identical, up to
a Lorentz boost in the $(t,x_1)$ plane, to the spherical reductions of
the non-extremal M2-brane, M5-brane and D3-brane that also have a pp
wave propagating on the world volume.  The Lorentz boost that relates
the two cases becomes an infinite boost in the extremal limit
$\mu_2\rightarrow \infty$.  This explains why the spherical reductions
give two distinct types of metric in the extremal cases, namely AdS if
there is no pp-wave, and the generalised Kaigorodov metric if there is
a pp-wave propagating on the original $p$-brane world-volume.

   In the non-extremal case, we have seen from the above discussion
that there is just the one type of metric to consider after spherical
reduction, namely (\ref{nonextnowavemet}), regardless of whether or
not there is a pp-wave propagating on the original $p$-brane.

Thus the general class of $D=n+3$ dimensional metrics that arises by
the spherical dimensional reduction of non-extremal boosted $p$-branes
is included in the class of metrics
\be
ds^2 = -e^{2a\rho+2f}\, dt^2 + e^{2b\rho}\, (dx + e^{(a-b)\rho}\,
dt)^2 + e^{2c\rho}\, dy^i\, dy^i + e^{-2f}\, d\rho^2\ ,\label{nonextk}
\ee
where $a$, $b$ and $c$ are constants, and the function $f$ is given by
\be
e^{2f} = 1 -k\, e^{-(a-b)\rho}\ .
\ee
(In obtaining the metric form (\ref{nonextk}) from (\ref{nonextmet}),
we have performed coordinate transformations that would not be valid
globally if $x_1$ were a periodic coordinate.  For the present
purposes we are principally concerned with local properties of the
metrics, for which this point is not essential.  If $x_1$ is
non-compact, ranging over the entire real line, the transformations
are in any case globally valid.)
 
     We find that (\ref{nonextk}) is an
dimensional Einstein metric if the constants $a$, $b$ and $c$
are given, as previously in the generalised Kaigorodov metrics in
(\ref{kaigconsts}), by
\be
a= (n+4)\, L\ ,\qquad b= -n\, L\ ,\qquad c=2 L\ ,\label{nonextconsts}
\ee
where again the cosmological constant $\Lambda$ is related to $L$ by
$L=\ft12\sqrt{-\Lambda/(n+2)}$.   Note that the Einstein metrics can
then be written in the form
\be
ds^2 =e^{-2nL\rho}\, dx^2+  e^{4L\rho}\, 
(2\, dx\, dt + k\, dt^2 + dy^i\, dy^i) + 
(1-k\, e^{-2(n+2)L\rho})^{-1}\, d\rho^2\ .
\label{nonextm}
\ee

    In the natural orthonormal basis $e^0 = e^{a\rho+f}\, dt$,
$e^1=e^{b\rho}\, (dx + e^{(a-b)\rho}\, dt)$, $e^2= e^{-f}\, d\rho$,
$e^i = e^{c\rho}\, dy^i$, we find that with the constants $a$, $b$ and
$c$ taking their Einstein-metric values (\ref{nonextconsts}), the
curvature 2-forms can be written in terms of the Weyl 2-forms $C_{ab}$
as $\Theta_{ab} = -4 L^2\, \eta_{ac}\, \eta_{bd}\, e^c\wedge e^d +
C_{ab}$, where $C_{ab}=C_{ab}^{\rm Kiag.} + \wtd C_{ab}$, with 
$C_{ab}^{\rm Kaig.}$ being the Weyl 2-forms for the Kaigorodov metric, as 
given by (\ref{weyltensor}), and 
\bea 
&&\wtd C_{01} = 2n \a\, e^0\wedge e^1 \ ,
\qquad \wtd C_{02} = 2n\, \a\, e^0\wedge e^2 + 2n(n+2)L^2\, (e^f-1)\,
    e^1\wedge e^2 ,\nn\\
&&\wtd C_{0i} = -4\a\, e^0\wedge e^i-2(n+2)L^2\, (e^f-1)\, e^1\wedge e^i
   \ ,\nn\\
&&\wtd C_{12} = 2n(n+1)\,\a\, e^1\wedge e^2 +2n(n+2) L^2\, (e^f-1)\, 
e^0\wedge e^2,\nn\\
&&\wtd C_{1i} = -2n\,\a\, e^1\wedge e^i-2(n+2) L^2\, (e^f-1)\, e^0\wedge e^i
\ ,\nn\\
&&\wtd C_{2i} = -2n\a\, e^2\wedge e^i\ ,
\qquad \wtd C_{ij} = 4\a\, e^i\wedge e^j\ ,\label{chnweyl}
\eea
where $\a= k\, L^2\, e^{-2(n+2)\rho}$.  Note that with the generalisations of
the Carter-Novotn\'y-Horsk\'y metrics written in the form (\ref{nonextm}), 
we regain the generalisations (\ref{kmet}) of the Kaigorodov metric simply
by setting $k=0$.  It can be seen from (\ref{chnweyl}) that the Weyl
tensor reduces to (\ref{weyltensor}) in this limit.
 
   With the metric written in the form (\ref{nonextm}), it is easy to
write down the Killing vectors. First of all, there are manifest shift
symmetries for all the coordinates $x$, $t$ and $y^i$.  In addition,
there are certain rotational symmetries in the $(t,x,y^i)$ hyperplane.
Thus there are in total $(\ft12n^2 +\ft32 n+2)$ Killing vectors, given
by
\bea
&&K_{(0)} = \fft{\del}{\del t}\ ,\qquad
K_{(x)} = \fft{\del}{\del x}\ ,\qquad
K_{(i)} = \fft{\del}{\del y^i}\ ,\nn\\
&&L_{(i)} = y^i\, \fft{\del}{\del t} - (x+k\, t)\, \fft{\del}{\del y^i}
\ ,\qquad
L_{(ij)} = y^i\, \fft{\del}{\del y^j} -y^j\, \fft{\del}{\del y^i} 
\ .\label{nonextkv}
\eea
Note that these non-extremal metrics have one less Killing vector than
the extremal generalised Kaigorodov metrics discussed previously,
since there is no longer a symmetry under which the coordinate $\rho$
is shifted while making compensating scale transformations of the
other coordinates.  Thus there is no longer an analogue of the $J$
Killing vector in (\ref{kvs}) in this case.  In the four-dimensional
case, there are now four Killing vectors.  Note that in all dimensions
the non-extremal Einstein metrics (\ref{nonextm}) are inhomogeneous.
This can be seen by calculating the curvature invariant
$R^{\mu\nu\rho\sigma}\, R_{\mu\nu\rho\sigma}$, which turns out to be
dependent on the coordinate $\rho$.

     As one would expect for spacetimes coming from the dimensional
reduction of non-extremal solutions, there are no Killing spinors in
the generalised Carter-Novotn\'y-Horsk\'y metrics.

    The four-dimensional case ($n=1)$ of the above metrics corresponds
to a previously-encountered solution.  In general, for arbitrary $n$,
let us define a new radial coordinate $R$, related to $\rho$ by
\be
e^{(n+2)L\rho} = \sqrt{k}\, \cosh((n+2)L R)\ .
\ee
In terms of $R$, the metrics (\ref{nonextm}) become
\bea
ds^2 &=& \Big(\sqrt k\, \cosh((n+2)L R) \Big)^{4/(n+2)}\, \Big[
k^{-1}\, (dx+ k dt)^2 - k^{-1}\, \tanh^2((n+2)L R)\,
dx^2\nn\\
&&+ dy^i\, dy^i \Big] + dR^2\ .
\eea
It is now easily seen that after simple coordinate transformations,
this metric in the four-dimensional case $n=1$ becomes equivalent to
the metric given in (13.32) of \cite{grbook}, which was found in this
form by Novotn\'y and Horsk\'y \cite{novhor}.  It is a special
case  of a general class of four-dimensional Einstein metrics found by
Carter in \cite{carter}.  

   In three dimensions, the Carter-Novotn\'y-Horsk\'y metric becomes
simply $ds^2 = dx^2 + e^{4L\rho}\, (2dx\, dt + k\, dt^2) + (1-k\,
e^{-4L\rho})^{-1}\, d\rho^2$.  This can be seen to be equivalent to
the BTZ black-hole metric described in \cite{btz}.

\end{document}